\documentclass[preprint]{aastex}

\usepackage{amsmath,amsfonts}

\newcommand{\rhoth}{\rho_\text{th}}
\newcommand{\lambdac}{\lambda_\mathrm{c}}
\newcommand{\lambdamin}{{\lambda_\text{min}}}
\newcommand{\lambdamax}{{\lambda_\text{max}}}
\newcommand{\Tmin}{{T_\text{min}}}
\newcommand{\Tmax}{{T_\text{max}}}

\usepackage[utf8]{inputenc}
\usepackage{natbib}
\usepackage{threeparttable}
\usepackage{url}
\usepackage{lscape}
\usepackage{soul}
\usepackage{lineno}
\title{Fermi GBM Observations of LIGO Gravitational Wave Event GW150914}
\shortauthors{Connaughton et al.}

\author{V.~Connaughton\altaffilmark{*,1}, 
E.~Burns\altaffilmark{2}, 
A.~Goldstein\altaffilmark{+,3},
L.~Blackburn\altaffilmark{4,5}, 
M.~S.~Briggs\altaffilmark{6,7}, 
B.-B.~Zhang\altaffilmark{7,8},
J.~Camp\altaffilmark{9},
N.~Christensen\altaffilmark{10},
C.~M.~Hui\altaffilmark{3},
P.~Jenke\altaffilmark{7},
T.~Littenberg\altaffilmark{1},
J.~E.~McEnery\altaffilmark{9},
J.~Racusin\altaffilmark{9},
P.~Shawhan\altaffilmark{11},
L.~Singer\altaffilmark{+,9},
J.~Veitch\altaffilmark{12},
C.~A.~Wilson-Hodge\altaffilmark{3},
P.~N.~Bhat\altaffilmark{7},
E. Bissaldi\altaffilmark{13,14},
W.~Cleveland\altaffilmark{1},
G.~Fitzpatrick\altaffilmark{7},
M.~M.~Giles\altaffilmark{15},
M.~H.~Gibby\altaffilmark{15},
A.~von~Kienlin\altaffilmark{16},
R.~M.~Kippen\altaffilmark{17},
S.~McBreen\altaffilmark{18},
B.~Mailyan\altaffilmark{7},
C.~A.~Meegan\altaffilmark{7},
W.~S.~Paciesas\altaffilmark{1},
R.~D.~Preece\altaffilmark{6},
O.~J.~Roberts\altaffilmark{18},
L.~Sparke\altaffilmark{19}, 
M.~Stanbro\altaffilmark{6}, 
K.~Toelge\altaffilmark{14},
P.~Veres\altaffilmark{7}
}

\altaffiltext{*}{Email: valerie@nasa.gov}
\altaffiltext{+}{NASA Postdoctoral Fellow}
\altaffiltext{1}{Universities Space Research Association, 320 Sparkman Dr. Huntsville, AL 35806, USA}
\altaffiltext{2}{Physics Dept, University of Alabama in Huntsville, 320 Sparkman Dr., Huntsville, AL 35805, USA}
\altaffiltext{3}{Astrophysics Office, ZP12, NASA/Marshall Space Flight Center, Huntsville, AL 35812, USA}
\altaffiltext{4}{Harvard-Smithsonian Center for Astrophysics, 60 Garden St, Cambridge, MA 02138, USA}
\altaffiltext{5}{LIGO, Massachusetts Institute of Technology, Cambridge, MA 02139, USA}
\altaffiltext{6}{Dept. of Space Science, University of Alabama in Huntsville, 320 Sparkman Dr., Huntsville, AL 35805, USA}
\altaffiltext{7}{CSPAR, University of Alabama in Huntsville, 320 Sparkman Dr., Huntsville, AL 35805, USA}
\altaffiltext{8}{Instituto de Astrof\'isica de Andaluc\'a (IAA-CSIC), P.O. Box 03004, E-18080 Granada, Spain}
\altaffiltext{9}{NASA Goddard Space Flight Center, Greenbelt, MD 20771, USA}
\altaffiltext{10}{Physics and Astronomy, Carleton College, MN, USA 55057}
\altaffiltext{11}{Department of Physics, University of Maryland, College Park, MD, USA 20742}
\altaffiltext{12}{University of Birmingham, Birmingham B15 2TT, United Kingdom}
\altaffiltext{13}{Istituto Nazionale di Fisica Nucleare, Sezione di Bari, 70126 Bari, Italy}
\altaffiltext{14}{Dipartimento di Fisica, Politecnico di Bari, 70125, Bari, Italy}
\altaffiltext{15}{Jacobs Technology, Inc., Huntsville, AL, USA}
\altaffiltext{16}{Max-Planck-Institut f\"ur extraterrestrische Physik, Giessenbachstrasse 1, 85748 Garching, Germany}
\altaffiltext{17}{Los Alamos National Laboratory, NM 87545, USA}
\altaffiltext{18}{School of Physics, University College Dublin, Belfield, Stillorgan Road, Dublin 4, Ireland}
\altaffiltext{19}{NASA Headquarters, Washington DC, USA}

\date{}

\newcommand{\fermi}{{\it Fermi }}
\newcommand{\Fermi}{{\it Fermi}}

\begin{abstract}
With an instantaneous view of 70\% of the sky,
the \fermi Gamma-ray Burst Monitor (GBM)
is an excellent partner in the search for electromagnetic counterparts to gravitational wave (GW) events.
GBM observations at the time of the Laser Interferometer Gravitational-wave Observatory (LIGO)
event GW150914 reveal the presence of 
a weak transient above 50 keV, 0.4~s after the GW event,
with a false alarm probability of 0.0022 (2.9$\sigma$).
This weak transient lasting 1 s was not detected
by any other instrument and does not appear connected with other previously known astrophysical, 
solar, terrestrial, or magnetospheric activity.
Its localization is
ill-constrained but consistent with the direction of GW150914. 
The duration and spectrum of the transient event
are consistent with a weak short Gamma-Ray Burst arriving at a large angle to the direction
in which \fermi was pointing, where the GBM detector response is not optimal.
If the GBM transient is associated with GW150914, this 
electromagnetic signal from a stellar mass black hole binary merger is unexpected.
We calculate a luminosity in hard X-ray emission between 1~keV and 10~MeV of 
 $1.8^{+1.5}_{-1.0} \times 10^{49}$~erg~s$^{-1}$.
Future joint observations of GW 
events by LIGO/Virgo and \fermi GBM could reveal whether the weak transient reported
here is a plausible counterpart to GW150914 or a chance coincidence, and will
further probe the connection between compact binary mergers and short Gamma-Ray Bursts.
\end{abstract}

\begin{document}
\maketitle

\section{Introduction\label{sec:intro}}
The  Gamma-ray Burst Monitor (GBM) on the \fermi Gamma-ray Space Telescope
is an all-sky hard-X-ray monitor that is ideally suited to detect rare and unpredictable transient
events.   Since the launch of \fermi in June 2008, GBM has triggered on-board nearly 5000 times
in response to short-lived impulsive bursts of photons lasting from under a millisecond to hundreds of seconds.
This collection of triggered events\footnote{\url{http://heasarc.gsfc.nasa.gov/W3Browse/fermi/fermigtrig.html}}
includes nearly 1800 Gamma-Ray Bursts (GRBs; \cite{vonkienlin2014}), 1100 solar flares, 
200 bursts from 9 separate magnetars, and
over 600 Terrestrial Gamma-ray Flashes (TGFs).
Dedicated offline searches over
all or parts of the mission have yielded over 200 additional magnetar bursts \citep{collazzi2015}, 
thousands of additional TGFs\footnote{\url{http://fermi.gsfc.nasa.gov/ssc/data/access/gbm/tgf/}} \citep{briggs2013},
nearly 700 type I thermonuclear bursts from galactic binary systems \citep{jenke2016},
non-impulsive steady or variable
emission from over 100 mostly galactic sources
 \citep{colleen2012}\footnote{\url{http://heastro.phys.lsu.edu/gbm/}},
and pulsed emission from 35 accretion-powered galactic
binary systems\footnote{\url{http://gammaray.nsstc.nasa.gov/gbm/science/pulsars.html}}.

Detection of gravitational waves (GW) reported by 
the Laser Interferometer Gravitational-wave Observatory (LIGO; \cite{aligo}) and the Virgo experiment \citep{avirgo}
has been eagerly anticipated.
LIGO and Virgo are sensitive to the GW produced by the mergers of stellar mass compact objects in a binary system as well as other sources.
The most promising electromagnetic counterpart to a compact binary merger involving a neutron star is a short GRB.
\cite{berger2014} gives a recent review of short GRBs and
\cite{fong2015} summarize what has been learned from observing their afterglows.
The joint GW-GRB detection rate is expected to be low given the collimation of the GRB emission
(both prompt and afterglow radiation) and the
detection horizons of LIGO and Virgo for these progenitors \citep{siellez2014}.  
Other electromagnetic counterparts have been suggested, 
notably optical/infrared transients from the decay of r-process isotopes produced in the ejecta
resulting from the
binary merger, kilonovae, also known as macronovae \citep{li1998,metzger2010,barnes2013,tanaka2013}. 
Only one such event has been reported \citep{tanvir_kilonova,berger_kilonova} with another 
candidate identified for a short burst with extended emission \citep{yang2015}.
Owing to their intrinsic faintness,
kilonovae are detected only nearby, an observational constraint aggravated by 
the need for short GRB localizations accurate enough to enable deep observations of the kilonova signal.
Unlike GRBs and their afterglow, however,
kilonovae are expected to be isotropic, so that mergers observed
off-axis by LIGO can still be probed for such a signature.   
GBM has detected 300 short GRBs since the launch of \Fermi, a rate of $\sim40$ per year \citep{vonkienlin2014}. 
An offline search of the GBM data yields an additional 35 short GRB candidates per year, most of them unverified 
by other instruments.
Validating these additional short GRB candidates and refining the search criteria will allow the GBM team to deploy an
efficient pipeline for the identification and communication in near real-time of sub-threshold short GRBs during 
upcoming observing runs of the LIGO and Virgo experiments.

Independently of this offline untargeted search of the GBM data, we 
developed a
targeted search and efficient data analysis pipelines to identify in the GBM data 
the electromagnetic counterparts to any candidate GW events.
We exercised and refined these pipelines
during Advanced LIGO's summer 2015 engineering runs in preparation for the first Advanced LIGO observing run (O1).
The search procedures and parameters, which we present here,
 were established {\it a priori} so that when LIGO began operations we could 
find candidate counterparts automatically, and establish their significance.
Joint localization of these events
may improve the localizations done separately,
which can assist follow-up observers with pointed instruments in identifying the host galaxy and thus the redshift of the source.  
On 2015 September 16, the LIGO and Virgo collaborations reported that a candidate event, G184098, had
 been identified in data recorded on September 14 \citep{gcn_ligo0,abbott2016}\footnote{Information about GW event
 candidates and follow-up observations was exchanged in Gamma-ray Coordinates Network (GCN) Notices and Circulars which initially were restricted to groups which had established agreements with LIGO and Virgo.  The Circulars regarding
 G184098 were added to the public archive when the details of GW150914 were published.}.
The candidate was subsequently characterized as being consistent with a signal from the
 merger of a stellar mass black hole binary system \citep{gcn_ligo1},
with a false alarm rate of less than one per century \citep{gcn_ligo2}, and was announced publicly in \cite{abbott2016} as GW event GW150914.
Although there are no predictions or well-established mechanisms for detectable EM emission from stellar mass binary black hole mergers 
to guide a search for counterparts in the GBM data, we carried out a methodical search around the time and sky location of the event GW150914,
which we report in the following section.

\section{GBM Observations of GW150914\label{sec:gbm}}
GBM consists of 12 
Thallium-doped Sodium Iodide (NaI) detectors with a diameter of 12.7~cm and a thickness of 1.27~cm and two Bismuth Germanate (BGO) detectors
 with a diameter and thickness of 12.7~cm \citep{meegan2009}.  
The NaI detectors are sensitive between 8~keV and 1~MeV and the BGO detectors extend the energy range to 40~MeV.  
The GBM flight software was designed
so that GBM can trigger on-board in response to impulsive events, if the count rates recorded
in two or more NaI detectors significantly exceed the background count rate on at least one time-scale (from 16~ms to 4.096~s) in
at least one of three energy ranges above 50 keV (50 -- 300~keV, $> 100$~keV, $> 300$~keV).
Strong background variations below 50 keV hinder the simple 
background fitting needed for automated operation on the spacecraft. On short time-scales, the variations are less significant and 
triggering can be enabled in the 25 -- 50 keV range on time-scales below 128~ms, resulting in the on-board detection of 200 magnetar bursts.  
GBM data can be probed at the longer time-scales and lower energy ranges
in offline searches dedicated to particular objects, including type I thermonuclear bursts \citep{jenke2016} and 
additional, weaker magnetar bursts \citep{collazzi2015}. The modification of the GBM flight software to include data from the BGO detectors 
in the 16 ms triggering window has made GBM very sensitive to spectrally harder events associated with the electric fields in
thunderstorms, Terrestrial Gamma-ray Flashes \citep{briggs2013}.

\begin{figure}
  \centering
   \includegraphics[width=7in]{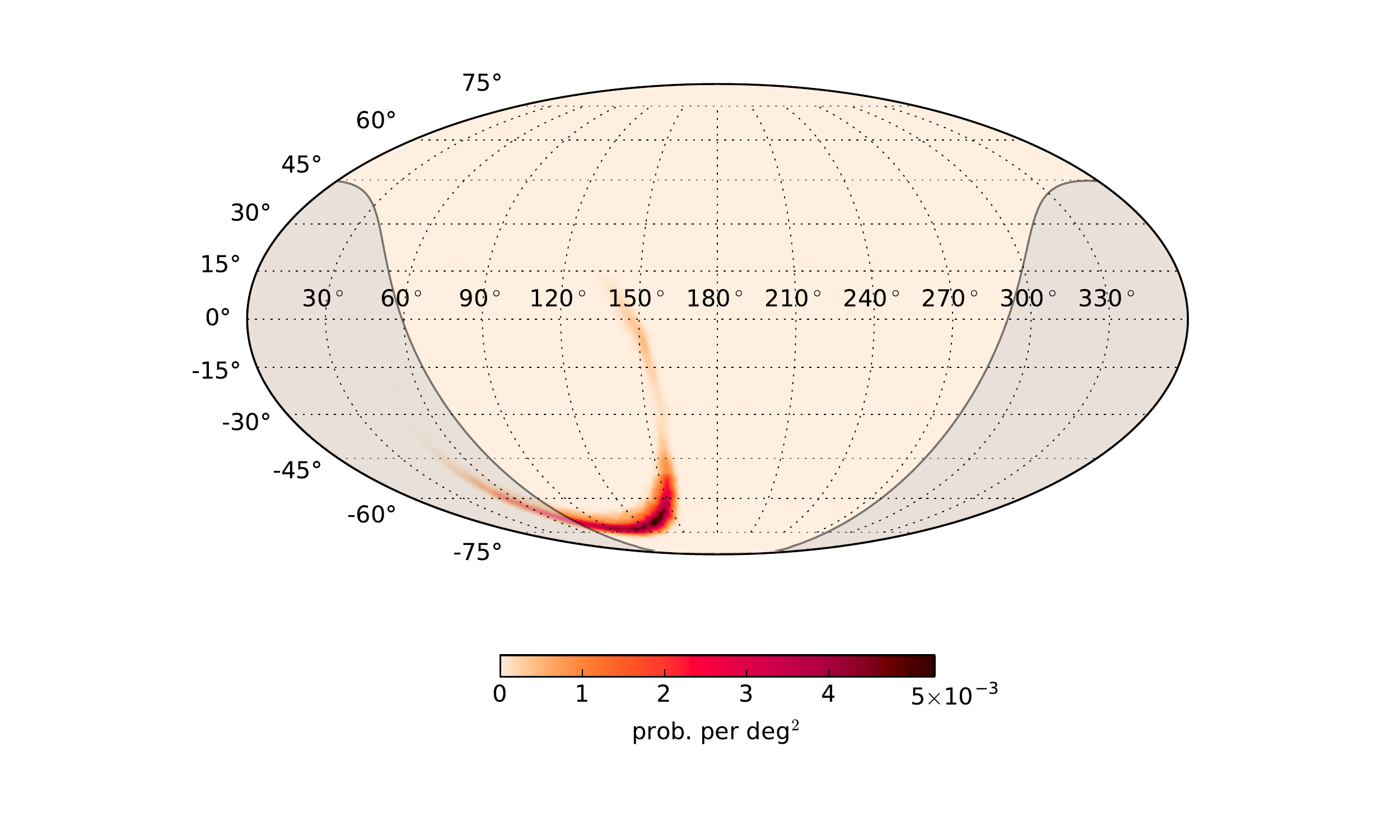}
   \caption{Localization map for GW150914, the GW event reported in \cite{abbott2016}. The grey shaded region indicates the
region of sky occulted to \fermi by the Earth at the time of GW150914.  The region not occulted by the Earth contains 75\% of the probability of the
localization map, with all but 6\% of the probability contained in the southern portion of the 
annulus.  The entire region was visible to \fermi
GBM 25 minutes after the GW event was detected. 
\label{fig:loc_ligo}}
\end{figure}

GBM has an instantaneous sky coverage of about 70\%, with the remainder blocked by the Earth.  GBM operates continuously except when
detector high voltages are turned off during passages of the \fermi spacecraft through regions of high particle 
precipitation activity in the South Atlantic Anomaly (SAA), $\sim$15\%
of the time depending on where \fermi is in the $\sim 50$-day precession cycle of its orbit.
GBM was recording data (i.e., not in the SAA)
continuously from nearly 2 hours before to over 7 hours
after the GW event.  Figure \ref{fig:loc_ligo} shows the LIGO sky map from \cite{abbott2016} with
the shaded region indicating the region of sky occulted to \fermi by the Earth at the time of detection of the GW event.  
GBM observed 75\% of the probability region in the
location map during the detection of GW150914, with the full region becoming visible 25 minutes later.

GBM did not record any on-board triggers around the time of the GW detection, at 09:50:45.391 UT on 2015 September 14.  
The triggers closest in time were from two events on 2015 September 14 
that are consistent with particle precipitation in or near the spacecraft, 
at 04:09:23 UT on entering the SAA,
and at 14:21:34 UT,  when \fermi was at high geomagnetic latitude, 
nearly 6 hours before and 4.5 hours after the GW event, respectively. 
GBM recorded triggers at similar points in the \fermi orbit on the preceding and following days, leaving no doubt as to their 
 magnetospheric origin.  These two triggered events were sufficiently far removed in time from GW150914 to ensure that GBM was operating
in a nominal configuration in which it could have triggered on significant transient sources above the on-board threshold.

\subsection{Detection and significance of weak, hard X-ray event GW150914-GBM\label{sec:discovery}}

An offline search of the GBM Continuous Time-Tagged Event (CTTE) data for
impulsive events too weak to trigger on-board \Fermi, or from a sky position unfavorable to 
the two-detector on-board triggering requirement, was implemented in 2015.  The main motivation for this offline search is
to increase the sensitivity of GBM to short GRBs during the period in which \Fermi, LIGO, and Virgo
operate jointly.  The offline search currently operates on CTTE data from the 12 NaI detectors over four energy bands 
(27 -- 540~keV, 50 -- 540~keV, 100 -- 540~keV, and 100 -- 980~keV) and 10 time-scales from
0.1 to 2.8 s. The detection threshold for each search algorithm
is set so that the joint chance probability of the signals in any detector combination exceeding background levels
above the lowest threshold level is $10^{-6}$ in one day. 
We estimate this improves GBM sensitivity to short GRBs by a factor of 2 -- 3 in burst count fluence
and the offline search
detection rate of 1 -- 4 candidate short GRBs per month is consistent with this estimate.  The offline search reports no candidates above
detection threshold on the day of the GW event\footnote{\url{http://gammaray.nsstc.nasa.gov/gbm/science/sgrb_search.html}}.

In addition to this undirected offline search, a 
targeted search of the GBM data was developed during S6, 
the last observing run of the previous
configuration of LIGO \citep{blackburn2015}.    
By searching both GW and GBM datasets, the significance of a sub-threshold signal in one can be strengthened 
by the detection of a signal in the other, provided the false positive rate of the joint search
is characterized and the detection levels in both instruments are selected accordingly. 
It is
estimated that the horizon of LIGO/Virgo can be boosted by 15 -- 20\% through this validation of sub-threshold candidates 
\citep{blackburn2015,kelley2013,kochanek1993}. 
The directed search of the GBM data is seeded with the time and (optionally) the sky location of any LIGO/Virgo candidate event. 
A coherent search over all GBM detectors (NaI and BGO) using the full instrument response at each sky position is performed 
over a user-specified time window, assuming one of three template source spectra,
revealing short-duration candidates typically between 0.256~s to 8~s in duration,
as described in Appendix \ref{app:search}. 
The candidates are ranked by a Bayesian likelihood statistic.

The model spectra for each tested source location are Band functions with three sets
of parameters spanning the range of astrophysical phenomena we expect to uncover.  
Emission from galactic transients, solar flares, and soft GRBs is expected
to favor a soft spectrum. Long GRBs are typically best fit with a moderate spectrum, and a hard spectrum is often preferred
for short GRBs.  
The values for the parameters of
the Band function \citep{band1993}, two power-law indices
and a peak energy, are those used in the standard GBM source localization process \citep{connaughton2015}:  
$\alpha$, $\beta$, $E_{\rm peak}$ =
(-1.9, -3.7, 70~keV),  (-1, -2.3, 230~keV), and (0, -1.5, 1~MeV), for the soft, moderate, and hard spectra,
respectively.  The response to each spectrum is evaluated over all sky locations with
an option to use a known source position as a prior in the evaluation of the likelihood.
Events, characterized by
their time and duration, are ranked by their likelihood ratios after
marginalizing over their unknown source amplitude, spectrum, and sky position.
We note that these spectral models are used as templates to identify candidates in the data, allowing 
a sky-position-dependent deconvolution of our data 
to evaluate the significance of any candidate across all detectors. No optimization of the models or of their parameters 
is performed. Because a trials factor is required for each template, we use only three
models, spanning a large parameter space from very soft to very hard, without any preconception about which type of event we
are seeking. Spectral analysis of any candidate is performed at a later stage (Section \ref{sec:spectrum}). 

We searched 30 seconds of GBM data before and after the LIGO coalescence time for a plausible counterpart with duration between 0.256~s and 8~s. 
The $\pm 30$~s interval we use was selected {\it a priori} and is roughly guided
by observation: if GRBs are related to compact binary mergers we expect the
impulsive gamma-ray emission to be close in time to the GW, suggesting an interval of just a few seconds for our search.
Precursors to short GRBs have, however, been observed earlier than $\sim$10~s prior to the main emission \citep{koshut1995,burlon2009,troja2010},
and may originate from a less collimated emission region that is observable even when the GRB jet is not along
the line of sight to the detector.

An all-sky search of the GBM data revealed two candidates below a threshold of $10^{-4}$~Hz chance probability. 
One transient, occurring at 09:50:56.8, 11~s after GW150914,
was visible only below 50 keV, favored the soft model spectrum, and lasted
2 seconds.  Using
the standard GBM localization procedure we found
a source position of RA, Dec =  267.7, -22.4 degrees, with a 68\%
statistical uncertainty region of radius $15^\circ$, and a systematic error of
around $3^\circ$ as described in \cite{connaughton2015}.
At a position in Galactic coordinates of l, b = 6.2, 2.4 degrees, the event is
compatible with an origin near the galactic center, well separated from and incompatible with
the LIGO localization region.  
It is typical of
the type of soft X-ray transient activity seen regularly in the GBM background data,
particularly from the galactic center region. We do not view this transient event
as being possibly related to GW150914 and we will not discuss it further.

\begin{figure}
  \centering{
   \includegraphics[width=5in]{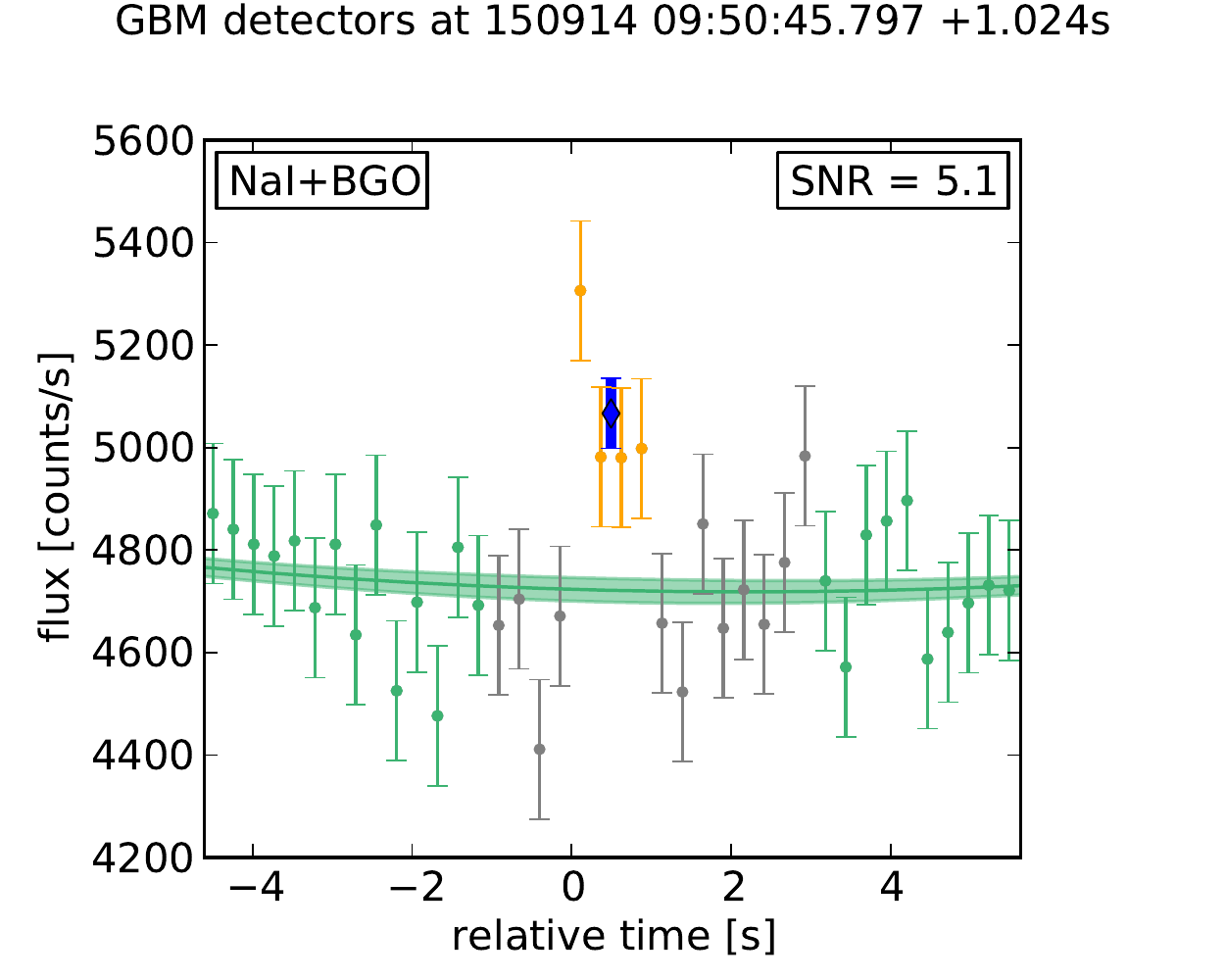}
   \caption{Model-dependent count rates detected as a function of time relative to the start of GW150914-GBM, $\sim$0.4~s 
after the GW event.  The raw count rates are weighted and summed to maximize signal-to-noise for a modeled source. 
CTIME time bins are 0.256~s wide. The teal data points are used in the background fit. 
The mustard points are the counts in the time period that shows significant emission, 
the grey points are outside this time period, and the blue point shows the 1.024 s average over the mustard points. 
For a single spectrum and sky location, detector counts for each
 energy channel are weighted according to the modeled rate and inverse
 noise variance due to background. The weighted counts from all NaI and BGO detectors are
 then summed to obtain a signal-to-noise optimized light curve for that model. Each
 model is also assigned a likelihood by the targeted search based on the foreground counts
 (in the region of time spanned by the mustard points), and this is used to marginalize the
 light curve over the unknown source location and spectrum.
\label{fig:discovery}}}
\end{figure}

The search also identified a hard transient which began at 09:50:45.8, about 0.4~s after the reported LIGO burst
 trigger time of 09:50:45.4, and lasted for about 1 second. The temporal offset of 0.4~s is much longer
than the light travel time of $2 - 45$~ms between \fermi and the LIGO detectors. 
The detector counts best matched those predicted from a hard model spectrum. 
We reported this event in \cite{gcn_gbm}; we henceforth call it GW150914-GBM. 
Figure \ref{fig:discovery} shows the model-dependent lightcurve of GW150914-GBM,
where the detector data have been summed using weights that maximize signal-to-noise for a given source model, and
 the unknown source model itself is weighted according to its likelihood in the data.

\subsection{The rate of detection of short hard transients in the GBM data\label{sec:far}}

\begin{figure}
  \centering{
   \includegraphics[width=5in]{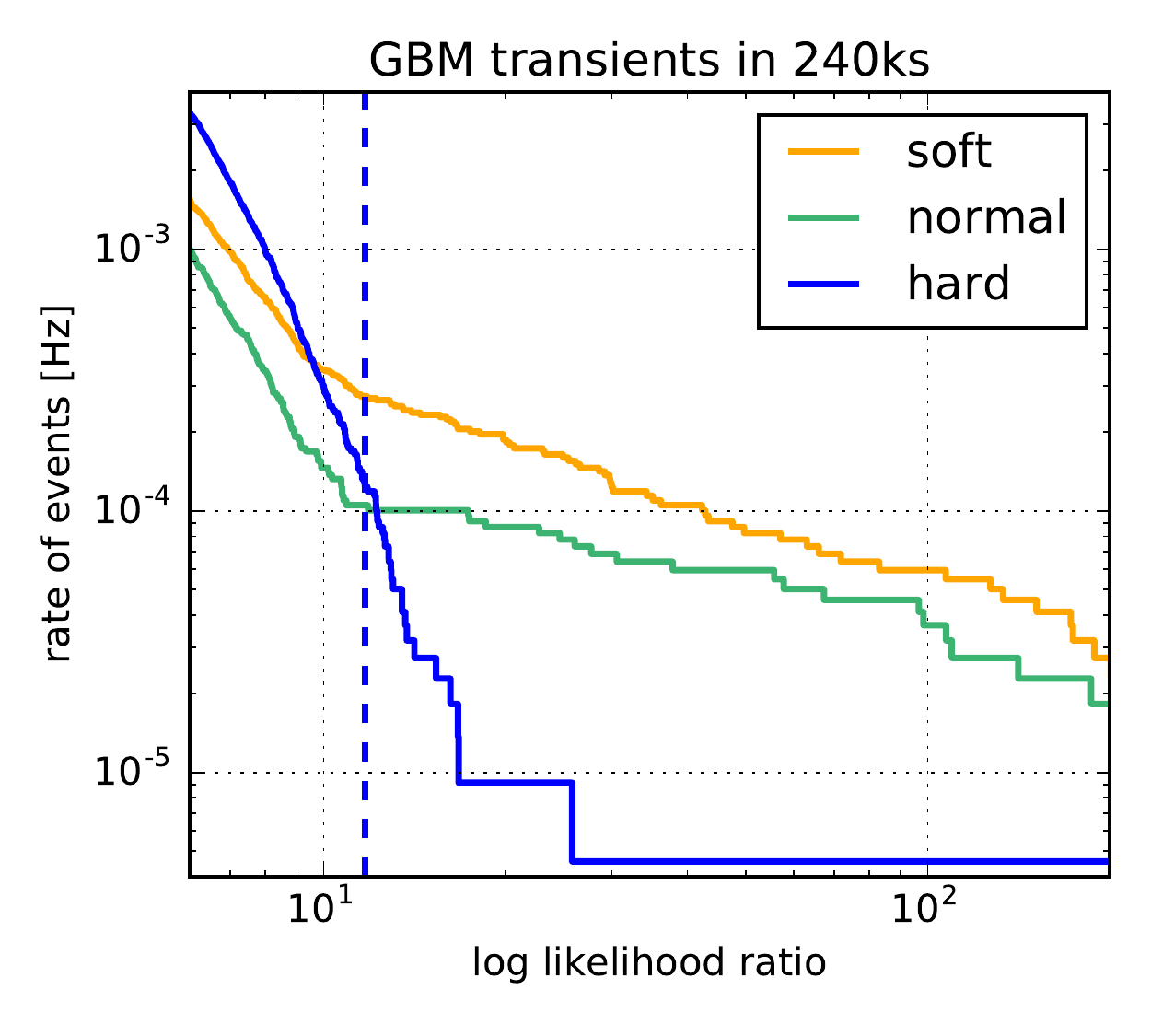}
   \caption{Distribution of transients identified by the targeted search pipeline in $\pm 120$~ks of GBM data 
surrounding GW150914. The events are between 0.256s and 8.192s in duration and sorted by best-fit
 spectral type. The dotted blue line marks the likelihood ratio assigned to nearby candidate GW150914-GBM, while
 the long-tail in the blue curve (hard spectrum) represents the single on-board triggered GRB in the
 data sample. The teal and mustard curves show the candidates that favor the other template spectra used
in the search.\label{fig:candidates}}}
\end{figure}

The association of a likelihood value with a false alarm rate (FAR) is based on an analysis
of two months of GBM data from 2009 -- 2010 \citep{blackburn2015}. The FAR for GW150914-GBM, $10^{-4}$~Hz, is very close to the 
reporting threshold for the search. The likelihood value for GW150914-GBM is much lower
than those obtained for two weak short GRBs detected by {\it Swift} that
did not cause an on-board GBM trigger but were found in a targeted search, and much higher than three
weak short GRBs that were undistinguishable above the background in the GBM data using our targeted search \citep{blackburn2015}.  
Because the likelihood value was so close to our reporting
threshold, we considered the possibility that the background count rates might be higher in 2015 than when the
search criteria and FAR were evaluated, implying a higher FAR than $10^{-4}$~Hz for GW150914-GBM.
We used our targeted search to examine 240~ks of GBM data from September 2015 with 218822.1~s of GBM live-time, excluding 
passages of \fermi through or close to the SAA where 
the detectors are turned off or count rate increases overwhelm any attempt to fit a reasonable background model.
We find 27 events above our threshold, 
 for a FAR of $1.2 \times 10^{-4}$~Hz, in agreement with the previously estimated value.
The distribution of events found in the 240~ks interval is shown in Figure~\ref{fig:candidates}.
This gives a 90\% upper limit on the expected background of hard transients of 35 in this much live-time, or 
$1.60 \times 10^{-4}$ Hz. 

We determine the significance of a GBM counterpart candidate by considering both its frequency of occurrence 
and its proximity to the GW trigger time.  
Our method, described in \cite{LIGO-T1500534}, and
attached as Appendix \ref{app:fap} to this work, 
allows us to account for all the search windows in the interval over which we
performed our search, while assigning larger significance to those events found closest
to the time of interest.  
This two-parameter ranking method frees us from having to choose a fixed search interval, and we can also limit the length of the search interval
to a value that is computationally reasonable.

With a false alarm rate of $1.60 \times 10^{-4}$ Hz for GW150914-GBM, which
begins 0.4~s after the time of the GW event, we calculate using Equation \ref{eq:fap} a post-trials
false alarm probability for GW150914-GBM,
P~=~$2 \times 3 \times 1.60 \times 10^{-4}$~Hz~$ \times$~0.4~s~$\times$~(1~+~$\ln$(30~s / 0.256~s))~=~0.0022 ($2.9 \sigma$),
where the logarithmic term accounts for the trials factor from multiple coincidence windows
and the factor of 2 accounts for the search window on either side of the GW time.
A trials factor of 3 is included to
account for the three spectral templates, 
which were treated as independent owing to their very different distributions. 

Our motivation for incorporating the temporal offset from the GW
 event into our likelihood ranking statistic is that we have a prior expectation that inspirals
 occur almost simultaneously with GRB production.  We do have a motivation for a search window that
 is long compared to the typical short GRB duration of 2~s so that our search is sensitive to precursors up to
 a few tens of seconds before the GW event.  
Most short GRBs do not, however, show precursor activity, so our {\it a priori} assumption is that
 a nearly simultaneous GBM transient is more likely to be associated with the GW event than one that is 10~s beforehand.
The false-alarm probability of coincidence scales very slowly with the selection of our window. 
For example if $\pm 60$~s were used instead, our calculated false-alarm probability would increase by only 12\%. 
Thus we believe our ranking strategy helps to reduce the
 dependence of calculated significance on specific tuning of these search parameters. 
If we assume, instead, a uniform probability across the 60~s window, we obtain a post-trials false-alarm probability of
$1-\exp(-60 \times 3 \times 1.60 \times 10^{-4}$)~=~0.028 ($1.9 \sigma$).

We now explore in detail whether the GBM data for
GW150914-GBM suggest an astrophysical origin and, if so, whether
the source is consistent with GW150914 or can be attributed to other causes.
We note that nothing in the following sections changes the FAR or the FAP that we present above.  
If further analysis of the 
data for GW150914-GBM suggested a non-physical source spectrum, or if the inferred brightness of the event proved incompatible with 
upper limits set by complementary observations, then this would lend support to a non-astrophysical nature for the
event, but it would not change the FAR  of the event or increase the probability that it occurred so close to
GW150914 by chance.  Similarly, if the search technique we developed proved inefficient, we could in principle 
improve our ability to discern real events and reject false ones, 
obtaining a lower FAR for a source associated with a given likelihood value. While we do not rule out future improvements based
on our experience during O1, we do not attempt here to improve our search {\it a posteriori}.
 
\section{Characteristics of GW150914-GBM\label{sec:character}}

Each GBM detector provides a different observational perspective. 
The relative rates in the NaI detectors 
establish the arrival direction of a source.
From the distribution of counts as a function of energy, we 
infer something about its nature. 
In Appendix \ref{sec:lc} we show that the detector pattern of GW150914-GBM is unusual,
with all individual detector count rates being slightly above background, simultaneously.
We also show that the count spectrum from
the NaI detectors (summed) is consistent with the count spectrum from the BGO detectors (summed), indicating a reasonable physical
spectrum that peaks in the BGO energy range.

\subsection{Localization\label{sec:localization}}

The angular response of the NaI detectors allows the reconstruction of the most likely
arrival direction of an impulsive event,
based on the differences in background-subtracted count rates 
recorded in 12 NaI detectors that have different sky orientations.  
A bright source is localized with a 68\%
confidence level statistical error of minimum  $1^\circ$ set by the resolution of a reference grid, 
and a systematic error that we have characterized in 
\cite{connaughton2015} as about $3 - 4^\circ$.
We can localize GW150914-GBM only roughly, as described in Appendix \ref{sec:loc},
to a region covering 3000 square degrees (68\% confidence level), with a most likely location of 
RA, Dec = 75, -73 degrees.
The source direction is 
underneath the spacecraft, at an angle of $163^\circ$ to the spacecraft pointing
direction, with 52\% of the probability region above the Earth limb, the
rest hidden by the Earth.

GBM was not designed to detect sources under the spacecraft, that have large angular offsets, $\theta$,  
to the spacecraft pointing direction. 
The pre-launch plan for
\fermi nominal operations was to observe at a $30^\circ$ angle from the local zenith, allowing the sky to drift across the field-of-view,
rocking the spacecraft north and south on alternate $\sim 90$-minute spacecraft orbits to achieve 
even sky coverage for the Large Area Telescope (LAT) survey of the high-energy sky. The GBM detectors were placed for
maximum sensitivity to sources in the LAT field-of-view ($\theta = $ 0 -- $\sim 65^\circ$), 
with good sensitivity out to $\theta < \sim 120^\circ$.  The 
Earth was expected to block the high $\theta$ regions, which are, by design, not well-viewed by the NaI detectors.    
The sky survey mode was changed after launch to alleviate the effect of higher-than-expected battery temperatures on the mission lifetime.  A
$50^\circ$ rocking profile was found to keep the batteries cooler and is now the nominal sky survey mode, 
with the result that GBM has more exposure to sky regions at high $\theta$ angles
than expected when deciding the detector placement.  
The combination of the declining sensitivity of the NaI detectors at large angles to the 
detector normals and the two-detector on-board trigger requirement results in
very few GRBs being detected with arrival directions at very high $\theta$. 

Of the 1776 GRBs listed in the Browse Table at the 
HEASARC\footnote{\url{http://heasarc.gsfc.nasa.gov/W3Browse/fermi/fermigtrig.html}}, only 67 occur
at a $\theta$ larger than $130^\circ$, and only 3 larger than $160^\circ$, none of the latter category short GRBs. One of the
GRBs detected beyond $160^\circ$, GRB130306A, 
was also detected by {\it Swift}. Because of the large uncertainty region associated with GW150914-GBM it is 
difficult to assess exactly how close its arrival direction is to that of GRB130306A,
but NaI~5 has the smallest angle to the source direction in both cases,
 and NaI~9 the largest.  GRB130306A showed roughly equal signals in all NaI detectors except NaI~10 and NaI~11, which were
fainter. GRB130306A
was a bright GRB, with a localization by GBM that was less than
$2^\circ$ from the {\it Swift} localization,
and a statistical uncertainty of $1^\circ$.  
This indicates that GBM is capable of localizing an event from an arrival direction beneath the spacecraft, 
from which nearly equal count rates are expected in most of the NaI detectors, if the event is bright enough.

\begin{figure}
   \center{
    \includegraphics[width=3.2in]{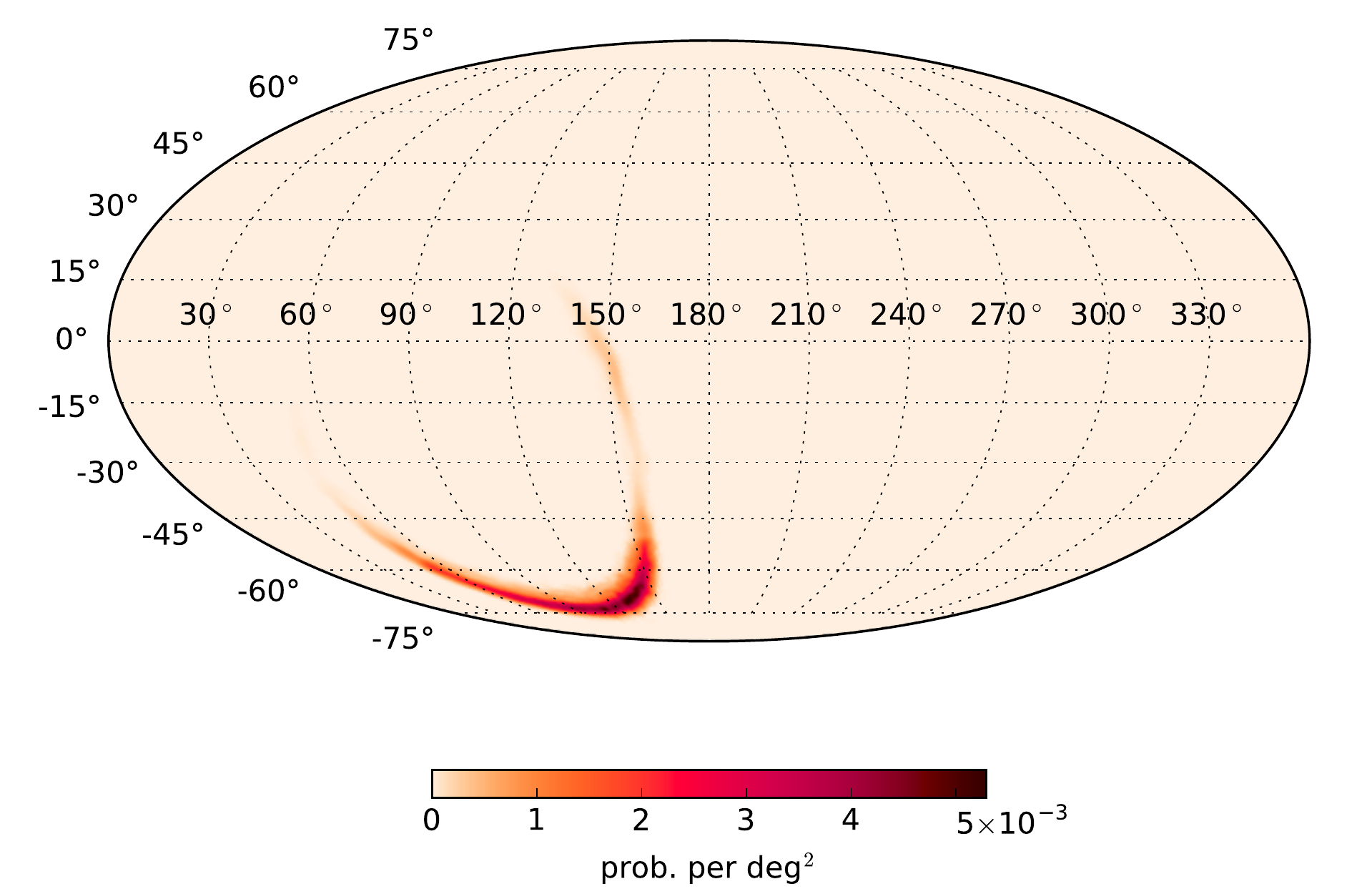}
    \includegraphics[width=3.2in]{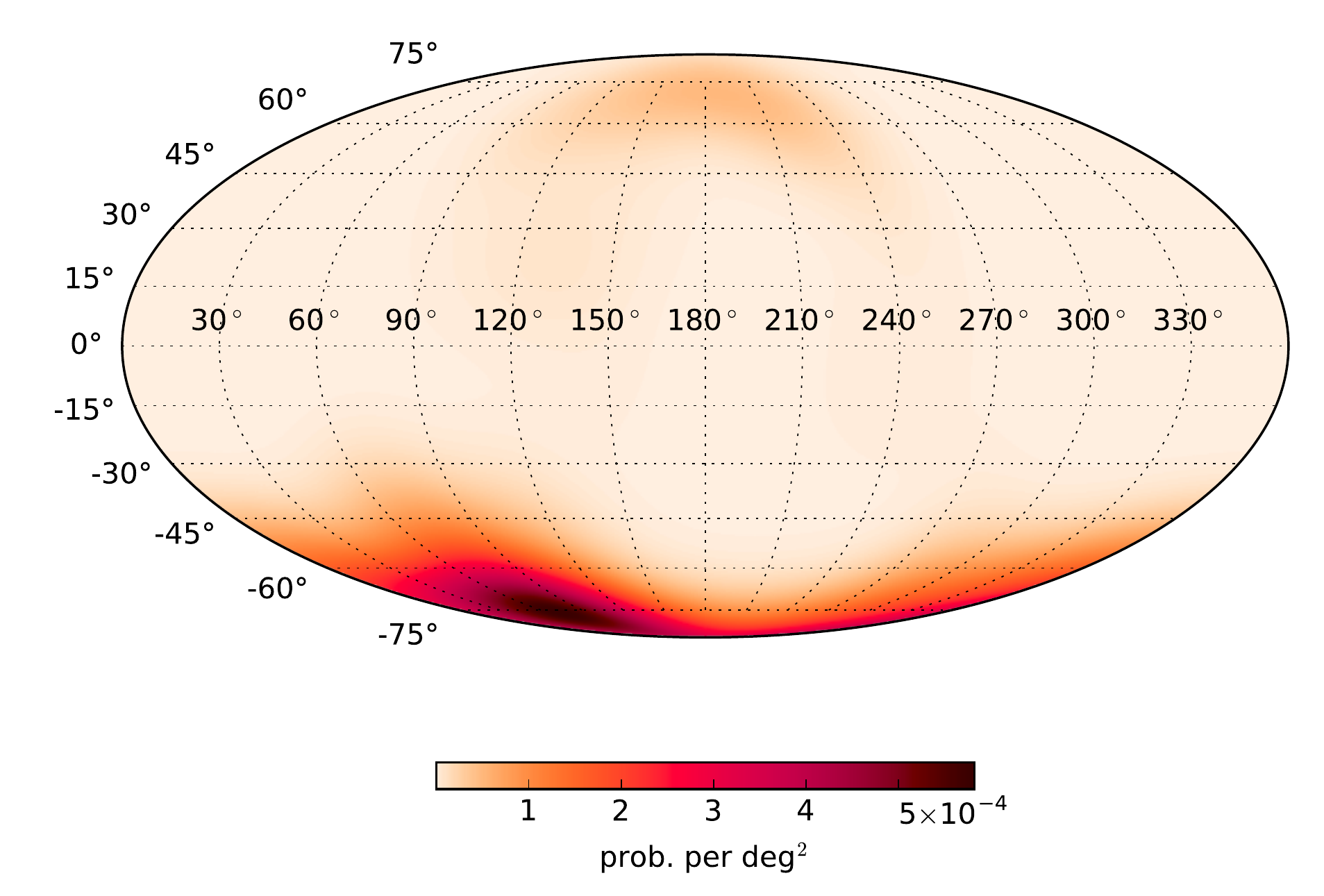}
    \includegraphics[width=3.2in]{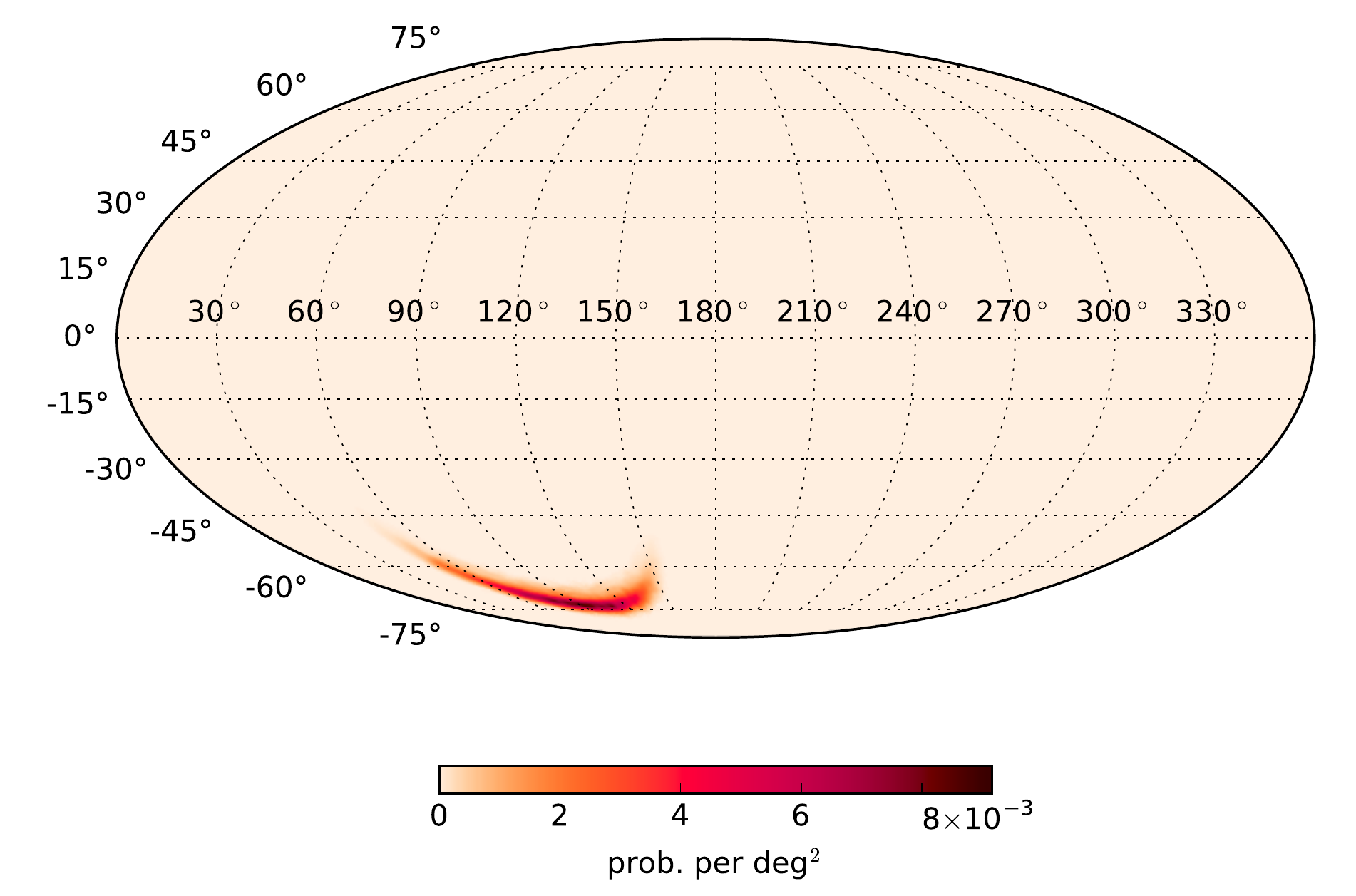}
    \includegraphics[width=3.2in]{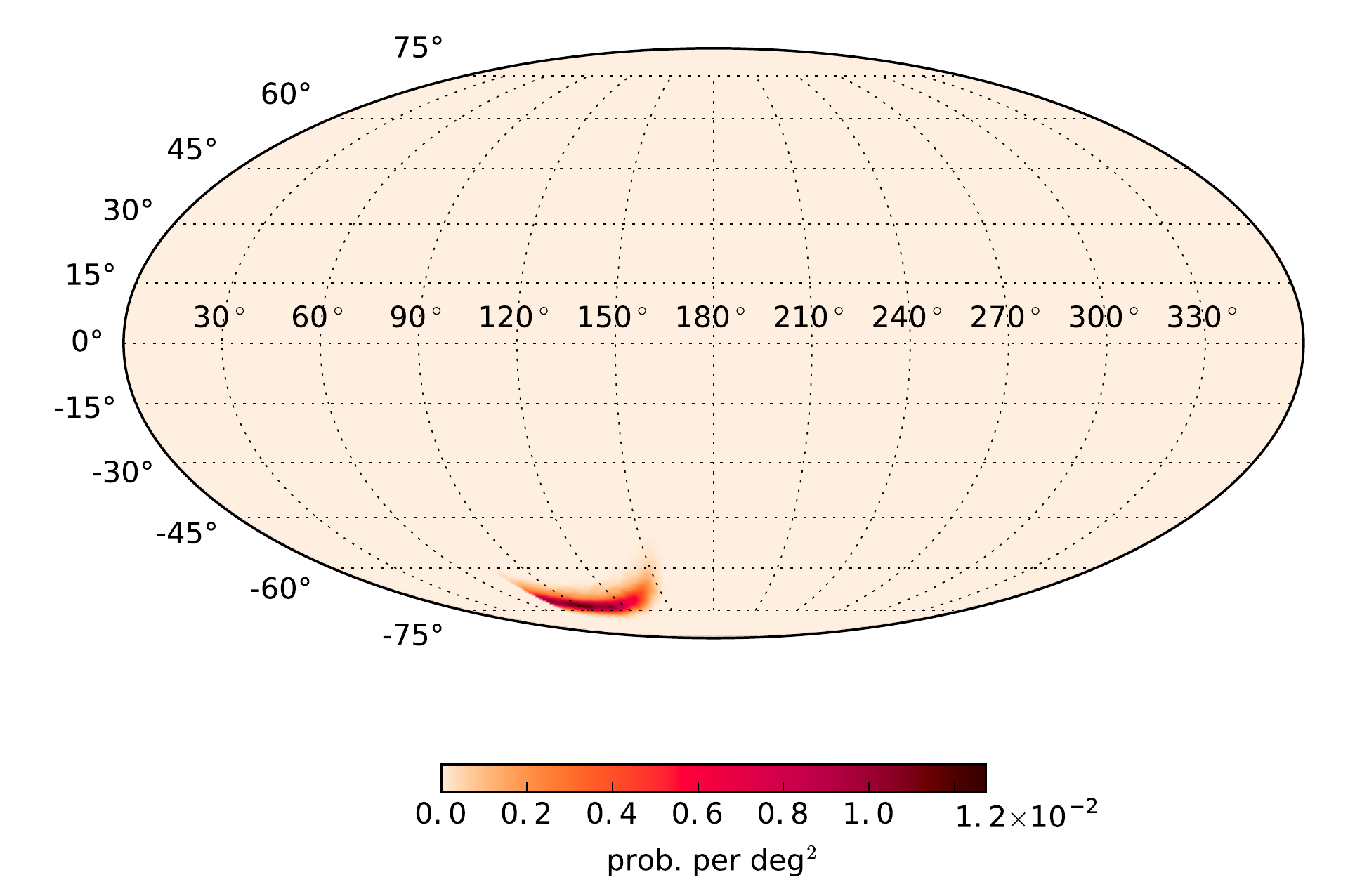}}
    \caption{The LIGO localization map (top left) can be combined
with the GBM localization map for GW150914-GBM (top right) assuming GW150914-GBM is associated
 with GW150914.  The combined map is shown (bottom left) with the sky region that is occulted to
\fermi removed in the bottom right plot.  The constraint from \fermi shrinks the 90\% confidence region for the LIGO localization
from 601 to 199 square degrees. \label{fig:loc_ligo_gbm}}
 \end{figure}

We find that the localization of GW150914-GBM is consistent with 
part of the LIGO localization annulus. 
If the transient event uncovered in the GBM data is associated with GW150914,
then the GBM probability map can be combined with the LIGO annulus
to shrink the 90\% confidence level LIGO localization by 2/3, as shown in
Figure \ref{fig:loc_ligo_gbm}.

\subsection{Energy spectrum of GW150914-GBM\label{sec:spectrum}}

The data for GW150914-GBM imply a weak but significant hard X-ray source with a spectrum that extends into the MeV range and a location
that is consistent with an arrival direction along the southern lobe of the sky map for GW150914.
Converting the observed counts in the GBM detectors to a source flux requires a deconvolution of the instrumental response 
with an assumed spectral model. 
We sample a range of arrival directions along the observed LIGO location arc, using the data and associated responses for the detectors
at each location that are most favorably oriented to the arrival direction. Table \ref{tab:arc} suggests that NaI 5 and BGO 0
are the most suitable detector set for all the locations along the arc. 
We use the rmfit spectral fitting package\footnote{\url{http://fermi.gsfc.nasa.gov/ssc/data/analysis/rmfit/}}, 
which takes a forward folding approach
to determine the parameters that best fit the data for any model, 
given the instrumental response.
The minimization routine producing the best fit parameters uses a likelihood-based fitting statistic, CSTAT. 

\begin{figure}
  \centering
    \includegraphics[width=5in]{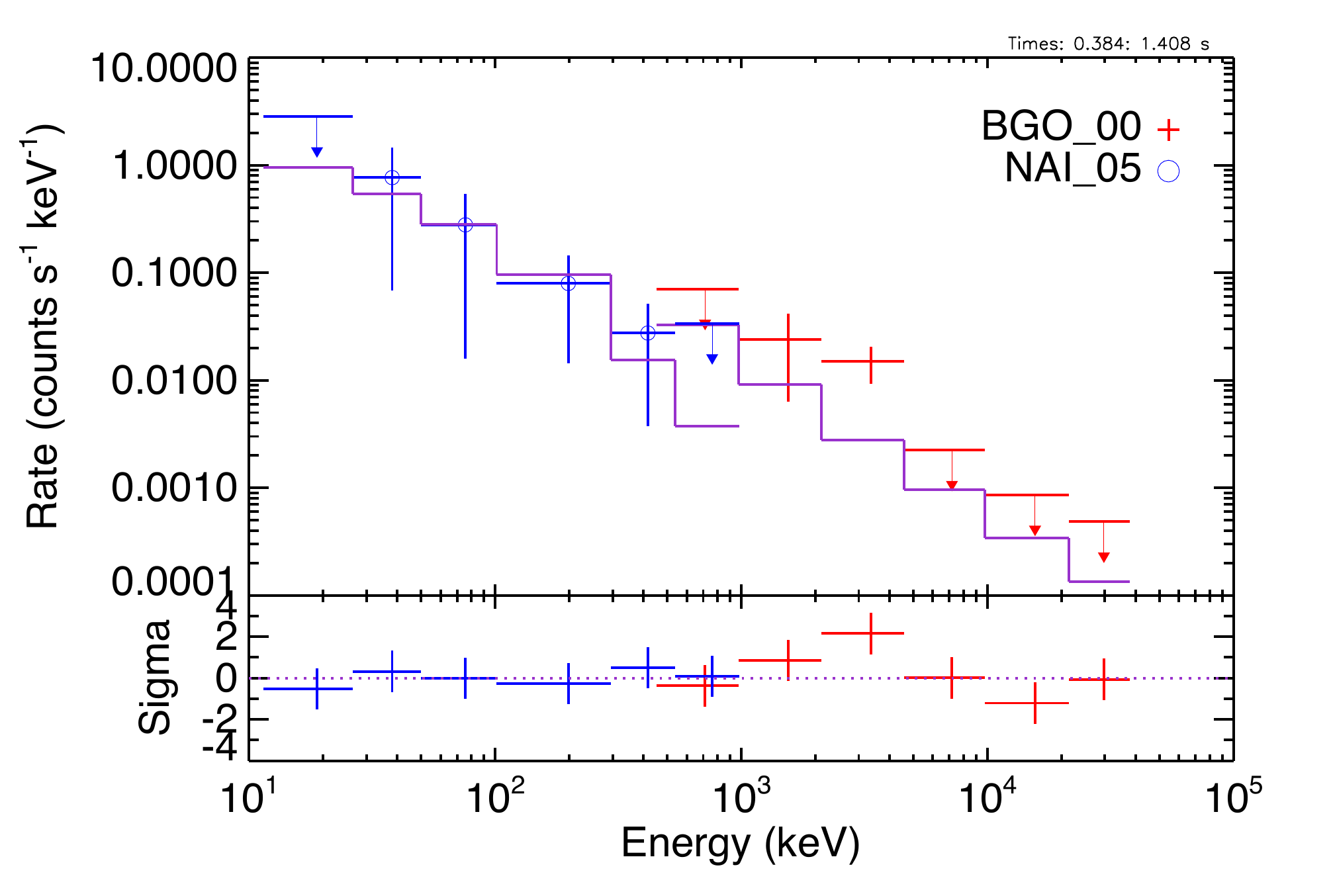}
   \caption{Power-law fit to the data 
from 0.384 to 1.408~s relative to the time of GW150914,
from NaI 5 (blue) and BGO 0 (red), corresponding to 
the high time bin in Figure \ref{fig:lc_tot}. 
The symbols show the data. The solid line shows the best-fit power-law model.
Residuals on the bottom panel show scatter but no systematic deviation.
We cannot use the first and last energy channels in either detector data type
(there are threshold effects and electronic overflow events), leaving 
the data from 12 energy channels included in the fit.  
\label{fig:spec}}
\end{figure}

Because the event is very weak, we do not attempt to fit the full-resolution data (128 energy channels).
Instead, we bin the CTTE data into the eight native CTIME energy bins,
and use the CTIME energy responses in our fits. In principle, binning in energy is unnecessary because a
likelihood-based statistic correctly accounts for low count rates in individual energy channels. In practice, the implementation
of CSTAT in our spectral fitting software neglects background fluctuations as a separate contribution to the uncertainty in the total
count rates in the GBM data, an effect that is
mitigated by rebinning the data prior to fitting. A consequence of this limitation of CSTAT is that the uncertainties on the parameters 
returned by the fits are almost certainly underestimated.  In the analysis that follows, we report 68\% statistical uncertainties, with
the caveat that the true uncertainties are probably higher.
GRB spectra are well represented by empirical functions with power-law components around a peak energy in the spectral energy distribution, 
$E_{\rm peak}$. The Band function is used when there are enough counts to constrain all parameters,
particularly the high-energy power-law index, $\beta$.  If $\beta$
is not constrained, a power-law fit with an exponential cut-off above $E_{\rm peak}$, called the Comptonized model, 
generally works well.  For the weakest bursts, or
when $E_{\rm peak}$ lies outside the energy range of the instrument,
a power-law fit is adequate and serves to provide an estimate of the
flux and fluence of the burst as long as the energy range over which the flux and fluence 
are calculated is not  extended outside the observation range.
We find that for all 11 positions along the LIGO arc,
a power-law fit to the data from GW150914-GBM can be constrained.  For one of the positions, we can also provide weak constraints
for a fit to the Comptonized model. 
Figure \ref{fig:spec} shows a representative count spectrum and power-law model fit 
to the data
from 0.384 to 1.408~s relative to the time of GW150914, with a deconvolution assuming the 
source lies near the central position of the southern arc.
For each of the 11 positions along the arc, we find the best-fit power-law index and associated amplitude. We use
these parameters to simulate each spectrum $10^4$ times, using the resulting distribution to estimate the 
uncertainties on the parameter values (68\% confidence level).  We also
sample the parameter distributions to calculate the fluence and its confidence region, weighting the sampling along the
arc according to the LIGO localization probability contained near each point on the arc.
We obtain a best-fit power-law index $-1.40^{+0.18}_{-0.24}$ and amplitude  $0.002^{+0.002}_{-0.001}$~photons~s$^{-1}$~cm$^{-2}$~keV$^{-1}$
over the LIGO localization arc, yielding a fluence between 10 and 1000~keV of $2.4^{+1.7}_{-1.0} \times 10^{-7}$~erg~cm$^{-2}$.

For a deconvolution assuming 
a source position at the northeastern tip of the southern lobe (entry 10 in Table \ref{tab:arc}), 
the Comptonized model converges to find a best fit 
$E_{\rm peak}$  of $3.5^{+2.3}_{-1.1}$~MeV with a power-law index below $E_{\rm peak}$ of
$-0.16^{+0.57}_{-0.50}$,
although this fit is not statistically preferred over the power-law fit. 
When simulating iterations of the 
burst to obtain 68\% confidence level uncertainties on the parameters, the fit failed about 50\% of the time. 
The fluence between 10 and 1000~keV obtained assuming a Comptonized model
for a source from this position 
is $2.8^{+1.0}_{-0.9} \times 10^{-7}$~erg~cm$^{-2}$.

The fit parameter values are typical for short GRBs, with power law indices of about -1.4 
found in cases where the GRB is too weak to constrain $E_{\rm peak}$,
and values for the Comptonized fit parameters that are not unusual for short GRBs \citep{gruber2014}.
A fluence of  $2.4 \times 10^{-7}$~erg~cm$^{-2}$ is nearly average for short GRBs, with 40\% of short GRBs detected by GBM weaker than this
value\footnote{\url{http://heasarc.gsfc.nasa.gov/W3Browse/fermi/fermigbrst.html}}. The least 
energetic short GRBs detected by GBM have a fluence an order of magnitude smaller than GW150914-GBM, implying that
if GW150914-GBM is a short GRB, then with a more favorable arrival direction, it would have caused an on-board trigger.
If GW150914-GBM is part of the short GRB population,
then its fluence is not atypical but its unfortunate arrival direction yields only a weak signal in GBM.  
Figure \ref{fig:spec} shows the model is a reasonable fit to the count spectrum even at low energies,
implying no paucity of counts at low energies in NaI 5, which is the only detector with a small enough viewing angle to the source position
to have any sensitivity below 50 keV. 

At a distance of $410^{+160}_{-180}$~Mpc implied by the GW observations \citep{abbott2016}, 
we obtain a source luminosity 
of  $1.8^{+1.5}_{-1.0} \times 10^{49}$~erg~s$^{-1}$ 
in the 1~keV -- 10~MeV energy range
that is standard for reporting such bolometric luminosities. 
The uncertainties reflect the range of possible distances to the progenitor,
uncertainties in the spectral fit parameters (using the power-law fits) and the range of arrival directions along the arc. 
This luminosity 
is an order of magnitude dimmer than the peak luminosities of the dimmest short GRBs in the sample analyzed by \cite{wanderman2015}.
     

\subsection{Other observations of GW150914-GBM\label{sec:other}}

Instruments other than GBM can also detect impulsive events in the hard X-ray energy range.  
No pointed instruments reported observations of GW150914, suggesting
they were not looking in that direction at the time of the GW event.

Upper limits to the emission from GW150914 from the non-detection
by instruments on-board the Astrorivelatore Gamma a Immagini Leggero (AGILE) close in time to the GW event
are reported by \cite{tavani2016}.  The  MicroCalorimeter (MCAL) had non-optimal exposure to the GW event, 
from which upper
limits to GW150914-GBM are calculated that are compatible with the GBM fluence. The other instruments
on-board AGILE observed most of the LIGO annulus hundreds of seconds either side of the GW event, 
but not at the time of the event.

The anti-coincidence shield (ACS) of the Spectrometer onboard INTEGRAL (SPI) has a large collection area above 80 keV with
an all-sky response that is not hindered by Earth occultation \citep{azk2003}.  
We looked for a signal in SPI-ACS at the time of GW150914-GBM and found no
excess above background\footnote{\url{http://isdc.unige.ch/~savchenk/spiacs-online/spiacs-ipnlc.pl}}.  The SPI-ACS team reported 
a fluence limit of $1.3 \times 10^{-7}$ erg cm$^{-2}$ in the 100~keV -- 100~MeV energy range based on a null detection over a 1~s period
\citep{gcn_integral}. 
Further analysis of the SPI-ACS data is reported in \cite{savchenko2016}.
They estimate a source signal between 5 and 15$\sigma$ above background 
should have been seen in the SPI-ACS data if
the source spectrum were similar to the template spectrum used in the discovery of GW150914-GBM.  
We note that the template spectrum was selected {\it a priori} and was not a fit to the data.     
A power law in energy with an index of about -1.4 was the only fit we could constrain
for a source at any position on the LIGO arc, with a Comptonized fit possible (but not statistically favored) for a
source at one of the positions we evaluated on the arc.  
Because power-law fits without a break are generally not physical representations of a source spectrum,   
a fluence calculation for the  
expected response in a detector with a different energy-dependent response than the instrument
in which the power-law fit was measured is not realistic.  
Instead,
\cite{savchenko2016} calculate the expected SPI-ACS signal
assuming various spectral shapes in an extrapolation from
the central value for the fluence obtained in fits to the GBM data. They report $3\sigma$ fluence
limits that are compatible with the $3\sigma$ fluence extrapolations obtained
in fits to the GBM data using the same assumed models.
 
\cite{azk2003} and \cite{savchenko2016} report that the sensitivity of SPI-ACS is a strong function of
the source energy spectrum and, to a lesser extent, the exposure of the detectors to the source location.  
In principle, non-detection by SPI-ACS can be used to further
constrain the spectrum and arrival direction of GW150914-GBM, or
indeed any event. 
If there is no allowed spectrum and location that can accommodate both an interpretation of the GBM excess as real and
the non-detection by the SPI-ACS, then we would conclude that a GBM signal was not astrophysical.
In the case of GW150914-GBM, the non-detection would appear to rule out a
 source with a spectrum as hard as the hard template spectrum used in the search pipeline, 
but not with the spectrum that was actually obtained in a fit to the data.
It might be constraining for a source represented by the Comptonized model with parameters 
obtained for a source at one position on the LIGO annulus, 
assuming the signal actually came from a source at another position on the annulus, to which SPI-ACS has optimum exposure. 
In practice, there are  
large uncertainties in both the spectrum and arrival direction of the source, and we
need a systematic study of the GBM and SPI-ACS sensitivities.
Because the SPI-ACS data are recorded with no energy resolution, as the sum of
nearly 100 detectors with different orientations, a joint fit to GBM and SPI-ACS data is not straightforward, using 
Monte Carlo simulations of the instrument response 
based on ground calibrations of the shield, with assumptions regarding the spectrum
of the background data recorded in orbit, and relying entirely on the spectral model and parameter values and
uncertainties obtained in fits to the GBM data.  In the case of GW150914-GBM, the GBM 
data alone cannot rule out a spectrum as hard as the model template in the discovery pipeine - 
the event is too weak to characterize the signal beyond a simple power-law fit and the BGO collection area above 10~MeV is too small for
a detectable signal in an event this weak. In principle, with
an understanding of systematic uncertainties in both the GBM fits and the cross-calibration of the two instruments,
 the flat response of the SPI-ACS above 10~MeV could constrain the spectral shape of GW150914-GBM.

Further investigation of the SPI-ACS detection sensitivity to GBM-detected GRBs as a function of the GRB
spectrum is ongoing, in order to evaluate the relative sensitivities of the two instruments to short GRBs, a study
involving both instrument teams that will include systematic effects that are neglected both here and in \cite{savchenko2016}.

\subsection{Possible origins for GW150914-GBM\label{sec:nature}}
The energy spectrum of GW150914-GBM is too hard for any of the galactic transient sources detected by GBM (bursts from magnetars, type I
thermonuclear X-ray bursts, or outbursts from accreting pulsars) and also too hard to be of solar origin.  Additionally, the Sun
was quiet around the time of the GW event detection. The localization (section \ref{sec:loc}) close
to the Earth's limb raises the question of whether GW150914-GBM comes from the Earth.  

Terrestrial Gamma-ray Flashes emit gamma rays
extending to at least 40 MeV.
TGFs are detected either as gamma rays produced
by electrons accelerated in electric fields in
thunderstorms, or as secondary electrons and positrons 
guided by the magnetic field line
that connects a thunderstorm to a gamma-ray detector.
Typical durations for the gamma-ray and electron events are several hundred $\mu$s and
several to tens of ms, respectively, much shorter than GW150914-GBM \citep{briggs2013}.
TGF gamma rays are detected by GBM when the source is within 800 km of the Fermi nadir;
the charged particle form can be detected from thousands of kilometers
from the source, but only when GBM is within the $\sim$100 km diameter beam
centered on the magnetic field line from the source \citep{dwyer2008,briggs2011,briggs2013}.
The World Wide Lightning Network (WWLLN; \cite{Rodger_WWLLN,WWLLN_Detection_efficiency}), a global network
of VLF radio receivers, virtually always finds clusters of lightning (i.e.,
thunderstorms) for GBM TGFs.
At the time of  GW150914-GBM WWLLN has no
lightning detections over $\pm$10 minutes within 800 km of the spacecraft
nadir nor at the two magnetic footprints, making it very unlikely that
there were TGF sources within GBM's detection range.

Another lightning detection network, GLD360 \citep{said2010,said2013}, reported a very high peak current lightning stroke at 
09:50:45.406 at  latitude 11.1685, longitude $-$3.2855 degrees.
At more than 4000 km from \Fermi, this is past the horizon so that
gamma rays would be blocked by the Earth.   The magnetic field line from this source
passes thousands of kilometers to the west of \Fermi, so if any
charged particles were emitted, they would not be transported to \Fermi.

At the time of the GW
event \fermi was at low geomagnetic latitude and was not near the SAA.
While we cannot exclude a magnetospheric origin for GW150914-GBM,
the observing conditions were not conducive to such an event, nor is the lightcurve typical of magnetospheric activity, 
which is usually manifested as longer and smoother (tens of seconds) bumps above background.

\subsection{Search for steady emission from known or unknown sources near the LIGO localization region\label{sec:occ}}

Using various search techniques,
we found (i) no evidence for long-term steady emission from the direction of GW150914-GBM, 
(ii) no evidence for contamination by known sources of hard X-ray emission  of any search for
emission related to GW150914-GBM,
and (iii) no evidence for non-impulsive emission related to the GW event in the days surrounding the event.

In addition to GBM's role as a powerful detector of transient, impulsive sources, 
the Earth Occultation technique (EOT) allows GBM to perform as an all-sky monitor of
sources emitting hard X rays at levels typically undetectable above the GBM background. 
This technique
involves modeling the GBM background count rates when a potential source of hard X rays sets or rises from behind the Earth.
Candidate sources are monitored\footnote{\url{http://heastro.phys.lsu.edu/gbm/}} with around 100 significantly
detected to date above 10 mCrab between 12 and 25~keV \citep{colleen2012}.  
Of the 246 sources that are monitored, five lie within $5^\circ$ of the LIGO localization region for GW150914:  
LMC~X-2, the flat spectrum radio quasar PKS~0601-70, the gamma-ray binary system 2FGL~J~1019.0-5856, and the accreting
X-ray binary pulsars GRO~J1008-57 and RX~J0520.5-6932 (which was detected in hard X-ray emission by {\it Swift} Burst Alert
Telescope (BAT) in 2013\footnote{\url{http://swift.gsfc.nasa.gov/results/transients/}}). 
Only GRO~J1008-57 has previously been detected by GBM through the EOT.
Both of the accreting pulsars lie within $3^\circ$ of the LIGO error region and have been detected in the past
through the GBM pulsar monitoring program, which
is more sensitive to pulsed emission than the EOT is to non-pulsed emission.
We looked for pulsed emission from these accreting
pulsars on 2015 September 14 and find they are not currently active. We also used a blind frequency search for pulsed emission 
from 24 positions along the Galactic plane and from the direction of the Small and Large Magellanic clouds.  We did not detect any signal
within or near the LIGO localization region.
In any search for long-lived emission in the days around the detection of the GW
event, we do not, therefore, expect contamination from known sources of hard X-ray emission above the GBM EOT and accreting
pulsar detection thresholds.

The daily sensitivity of the EOT
is about 100 mCrab. The EOT can resolve signals from sources $2^\circ$ apart.  We divided the
full LIGO arc into 34 resolvable positions (all but one along the southern lobe of the arc)
 and looked for mission-long activity from these positions, as well as
daily emission around the time of the GW event.  
We examined 3 years of data using the EOT, from 2013 January 1 through 2016 January 29. 
Long-term averages were consistent with no detections for the 12 -- 25, 25 -- 50, 50 -- 100, 100 -- 300, and 300 -- 500 keV energy bands. 
We also looked for emission on a daily time-scale for
the month of September 2015 without detecting any of the sources during the month surrounding the LIGO GW event time. 

The Earth occultation technique fails
to measure source fluxes if the angle between the tangent to the Earth’s limb and the spacecraft orbit normal, $\beta$, exceeds $66.5^\circ$.  
At grazing incidence, the Earth occultation transition becomes too extended in time ($>$20 s from 100\% -- 0\% atmospheric transmission),  
and at $\beta$ values beyond grazing incidence, the source is not occulted by the Earth at all.
This occurs at certain points in the $\sim 50$-day \fermi 
orbital precession cycle for high declination sources ($> \pm 40^\circ$) owing to the relative geometry of the source position and
the Fermi orbital inclination of $26.5^\circ$.  
Only 13 of the targets, with right ascensions from 48 -- 77$^\circ$, 
and the northern lobe position, had usable Earth occultation measurements spanning the time of the LIGO event. 
The remaining targets with right ascensions from 74 -- 155$^\circ$
 had no usable Earth occultation measurements from before the time of the LIGO event until 2 or more days after GW150914. Another way to look at
this is that these unocculted positions never set behind the Earth and
were observed by GBM with 85\% exposure, losing only the time that \fermi crossed through the SAA. 
For much of the LIGO arc during the days around the GW event detection, 
GBM was thus exceptionally sensitive to any impulsive emission that would have triggered the instrument.

If GW150914-GBM is related to the GW event,
and the localization is in the region of the LIGO arc with $\beta \sim 66.5^\circ$,  i.e. very close to being occulted by the Earth,
 then grazing Earth occultations could be responsible for a reduction of flux below 50 keV through atmospheric absorption (Figure \ref{fig:lc})
 and could potentially be used to further improve the source location.
Lower energy photons, e.g. 12 -- 25 keV can be fully blocked (0\% atmospheric transmission) before the 100 -- 300 keV band reaches 50\% transmission.
We cannot exclude the possibility that the spectral analysis (and thus the luminosity estimate) is affected by partial, energy-dependent 
atmospheric absorption of the signal, but the spectral deconvolution of the data from NaI 5 (section \ref{sec:spectrum})
does not suggest a deficit of counts below 50 keV relative
to the model. It is more likely 
that the source is not close to being occulted by the Earth and, instead, 
that the hard spectrum observed in most of the detectors is a mixture of intrinsic spectral hardness and
the large viewing angles to most of NaI detectors which lead to preferential detection of higher-energy photons 
and absorption of photons of lower energy in the instrument material behind the scintillator.

The LIGO localization arc for GW150914 became observable by the \fermi LAT
$\sim$4000~s after the GW event and a search for high-energy emission over time-scales comparable to our search in hard X rays with the EOT
is reported by \cite{omodei2016}.  A summary of observations of GW150914 is given in \cite{singer2016}.

\section{Discussion and outlook for joint LIGO-GBM science\label{sec:discussion}}

GBM observed over 75\% of the probability 
in the GW event sky location at the time of GW150914.  
A weak hard X-ray transient lasting around 1~s was detected above 50 keV 0.4~s after the GW event
using a technique developed to find short transients in the GBM data in coincidence with sub-threshold GW events.  
The chance probability of finding such an event within the time interval we searched is
0.2\% following the assumption, made {\it a priori}, that
the likelihood of a counterpart associated with the GW event is higher
for an event closer to the time of the GW event, and 2.8\% if we assume equal probability of association
across the 60~s search window. 
The GBM signal is localized to a region consistent with the LIGO sky map, with a large uncertainty on the location. 
If the transient event uncovered in the GBM data is associated with GW150914, then it is possible its origin 
under the \fermi spacecraft, combined with the weakness of the source, can account for the lack of confidence 
associated with the standard localization procedure applied to this event.
If we assume the LIGO and GBM events have a common origin, then combining the LIGO and
GBM localization maps reduces the LIGO localization area by 2/3.

The transient event cannot be attributed to other known astrophysical, solar, terrestrial, or magnetospheric activity.
The distribution of detected counts as a function of energy
appears reasonable among detectors across the energy range 50 keV -- 4.8 MeV. Spectral deconvolution
yields a fluence (68\% confidence level)
over the 1 s duration of $2.4^{+1.7}_{-1.0} \times 10^{-7}$~erg~cm$^{-2}$ between 10~keV and 1~MeV,
comparable to moderate intensity short GRBs on which GBM has triggered.  
This implies that with a more favorable arrival geometry, this event could have triggered GBM on-board
at the time of the GW detection, providing a real-time localization within seconds of the trigger.
A real-time electromagnetic counterpart to a GW event informs follow-up observers that an afterglow signal
may be detectable along the line of sight, and the GBM location helps in reducing
 the number of observations needed to cover the LIGO localization region.  

The collection area of SPI-ACS is a factor of $\sim 30 - 40$ times greater than that of a GBM BGO detector, but the
upper limits imposed by the non-detection of GW150914-GBM by SPI-ACS are close to the fluence values calculated for the GBM transient, 
despite the unfavorable source direction for GBM and the optimum source direction for SPI-ACS.  Because of uncertainties
in the energy spectrum and
location of GW150914-GBM, and because of instrumental and background systematic effects 
on the calculation of the fluence of GW150914-GBM, any tension between the GBM and INTEGRAL SPI-ACS observations 
will likely be resolved only with future
joint observations of GW events. 


The detection of an electromagnetic counterpart to a 
merger of stellar mass black holes would be a surprising event.  
\cite{kamble2013} explore possible weak
signatures to such mergers, with uncertainties surrounding
the formation of circumbinary disks and associated magnetic fields. 
Although circumbinary disks are expected to form around supermassive black holes \citep{mayer2007}, 
there is no such prediction for stellar mass systems.  Moreover, the GBM signal appears similar to a
short GRB, both in duration (less than 2~s), and in energy spectrum (peaked near an MeV).  
Models for short GRBs from compact binary progenitors always involve a neutron star,
with short GRBs more easily produced from two neutron stars, unless the 
black hole companion has a high initial spin \citep{giacomazzo2013}.
A luminosity 
of $1.8^{+1.5}_{-1.0} \times 10^{49}$~erg~s$^{-1}$
(between 1~keV and 10~MeV) 
for a short GRB, assuming the source 
 distance of $410^{+160}_{-180}$~Mpc implied by the GW observations \citep{abbott2016},
is an order of magnitude dimmer than the peak luminosities of the dimmest short GRBs in the sample analyzed by \cite{wanderman2015}. 
By another measure of brightness, the isotropic-equivalent energy release, also measured between 1~keV and 10~MeV, 
GW150914-GBM would also be dimmer than most short GRBs, but similar in magnitude to GRB050709 and GRB080905A, which were also nearby (z=0.161 and 
0.122, respectively), 
and an order of magnitude dimmer in isotropic-equivalent energy release than the next dimmest short GRB reported
in \cite{giacomazzo2013} and \cite{davanzo2014}. 
If GW150914-GBM is a short GRB, then it was detected only because it was nearby. Based on the population of
short GRBs with known redshifts, the contribution of such under-luminous events
to the overall short GRB population detected by GBM is negligible, unless they form a separate class of
nearby, sub-luminous events.

Our observation of GW150914-GBM has spurred investigations into 
complementary observations that may reveal afterglow signatures of such events \citep{yamazaki2016,morsony2016},
a possible mechanism to extract high-energy
emission from stellar mass black hole mergers \citep{zhang2016},
unusual environments for the black hole merger that may lead to sufficient surrounding material
to fuel the production of the GRB \citep{loeb2016,perna2016},
implications of our observation if the association between GW150914 and GW150914-GBM is real \citep{ellis2016,li2016},
as well as arguments against the association being real, based on the difficulties extracting enough energy
from the black hole merger \citep{lyutikov2016}. 

Further observations by LIGO and Virgo in coincidence with a detector sensitive to hard X-ray or gamma-ray 
transient events will determine whether short bursts of high-energy electromagnetic radiation accompany 
stellar mass black hole binary
mergers.  Because of the weakness of GW150914-GBM and its 
large localization uncertainty, chance coincidence may play a role in both  
the identification of GW150914-GBM as an astrophysical phenomenon and its association with the GW event,
even with the false alarm probability of 0.0022 that we calculate in section \ref{sec:far}. 
If the association is real, then the alignment of the merger axis with our line of sight may be attributed
in part to the greater sensitivity of
LIGO to on-axis events, but we would not expect most GW signals from BH mergers to be accompanied by the detection of
collimated electromagnetic transients.
Another possibility is that
the electromagnetic emission is not narrowly collimated and we can expect further joint detections of stellar mass black hole binary mergers and GRBs.    
This paradigm may be in tension with the non-detection of GW candidates in the last science
runs of the previous configuration of LIGO/Virgo, S6/VSR2\&3 \citep{ligo_s6_grbs}. 
None of the GRBs with known redshift detected during S6/VSR2\&3 was within the
BBH detection horizon ($\sim$100~Mpc).  It is possible, however, that some of the 90\% of GRBs with unknown redshifts
 were within the BBH horizon, which is, nonetheless, much closer than most short GRBs.

Analysis of the GBM data corresponding
to all sub-threshold GW events from the O1 initial science operation 
period of LIGO is in progress.  
We have developed
pipelines and data products to rapidly search the GBM data for counterparts to any GW events and communicate their localization
to electromagnetic observers within hours of the GW event (depending on data downlink from the \fermi spacecraft).  

Given the detection of GW150914 as a GW event from a stellar mass black hole binary system, then with all but
the most pessimistic predictions, the detection of the weaker 
GW signals from neutron star binary systems is 
expected no later than 2019, when LIGO/Virgo reach full sensitivity.  If this detection 
occurs during O2, the second observing run of LIGO and the
initial deployment of Virgo, expected later in 2016, our GBM-LIGO/Virgo pipelines are ready. 
Even if the association between GW150914-GBM and GW150914 is spurious, we expect to detect short GRBs from neutron star binary systems.
With its broad field-of-view and good sensitivity at the peak emission energies for short GRBs, 
\fermi GBM is an ideal partner in the search for electromagnetic signals in coincidence with gravitational wave detections.
Joint observations by \fermi and LIGO/Virgo will either confirm
 or exclude the connection between compact binary systems and short GRBs within a few years.

\begin{acknowledgments}
The authors acknowledge an ongoing collaborative effort and significant
contributions to this paper from the LIGO/Virgo Consortium.
They thank Marica Branchesi for her co-leadership of the LIGO/Virgo electromagnetic follow-up group, 
for providing prompt alerts and other information pertaining
to the GW event. 
The GBM project is supported by NASA.  Support for the German contribution to GBM was provided by the Bundesministerium f\"ur
Bildung und Forschung (BMBF) via the Deutsches
Zentrum f\"ur Luft und Raumfahrt (DLR) under contract number 50 QV 0301.
A.v.K. was supported by the Bundesministeriums f\"ur Wirtschaft und Technologie (BMWi) through DLR grant 50 OG 1101.
AG and LS are funded through the NASA Postdoctoral Fellowship Program.
OJR acknowledges support from Science Foundation Ireland under Grant No. 12/IP/1288.
JV was supported by STFC grant, ST/K005014/1.
NC acknowledges NSF grant, PHY-1505373.
The authors wish to thank the World Wide Lightning Location Network (\url{http://wwlln.net}), a collaboration among over 50 universities and institutions, for providing the lightning location data used in this paper. 
The authors are grateful to Sylvia Zhu for a critical reading of the manuscript and for her helpful suggestions, and also
appreciate the constructive feedback from three anonymous referees.
\end{acknowledgments}
 
\bibliographystyle{apj}

\hyphenation{Post-Script Sprin-ger}

\newpage
\appendix

\section{A targeted search of GBM data\label{app:search}}

The targeted search attempts to identify short-duration ($\sim$1~s)
excesses of counts recorded across the detectors that stand out over a
smoothly-varying background and that are consistent with a modeled point-source
contribution from an astrophysical event. The seeding is done in time and sky
position, where the seed time defines a limited ($\sim$minutes) period of time
to scan, and a seed sky position prior can be used to inform the model prior.

The short-duration excess of counts from an astrophysical event are
hypothesized to occur over a foreground interval [$t-T/2$, $t+T/2$]. Trial
foreground durations $T$ are spaced in powers of 2 between 0.256 and 8.192
seconds, and for a given duration, central times $t$ are chosen at 75\%
time-interval overlap. This choice approximately preserves signal-to-noise
mismatch across the search space. 
The technique was developed prior to the availability of
CTTE data, using CTIME data, which are natively
binned in 0.256 s accumulations with counts binned in 8 energy channels. 
The counts registered in the 14 GBM detectors and 8 energy channels are evaluated
independently for each detector-channel combination. For each short foreground
interval [$t-T/2$, $t+T/2$], we estimate the background rate at $t$ using a
polynomial fit to local data from [$t-10T$, $t+10T$] (minimum $\pm5$~s),
excluding time [$t-3T/2$, $t+5T/2$] around the foreground interval to avoid
bias from an on-source excess. The polynomial degree is determined by the
interval length to account for more complicated background variability over
longer intervals. It ranges from 2 (minimum) to 1+$0.5\log_2 T$.

A likelihood-ratio statistic is constructed for measured counts within this
foreground interval that compares the hypothesis that detector counts arise
from the expected background contribution plus a modeled signal (hypothesis
$H_1$) to the hypothesis that observed counts arise from variations in
estimated background rates alone (hyothesis $H_0$). The likelihood of observed
background-subtracted counts $\tilde{d_i} = d_i - \langle n_i \rangle$ to have
arisen solely from background fluctuations alone is,
\begin{equation}
P(d_i|H_0) =
    \prod_i{\frac{1}{\sqrt{2\pi}\sigma_{n_i}}
        \exp\left(-\frac{\tilde{d_i}^2}{2\sigma_{n_i}^2}\right)}
\end{equation}
where $i$ runs over all independent measurements from detector-channel
combinations (14 detectors, 8 channels) and $\sigma_{n_i}$ represents the
standard deviation for each measurement under a Gaussian approximation to the
Poisson process. If we include expected source contributions $r_is$ (source
amplitude $s$ subject to instrument response $r_i$) to each measurement from a
modeled source, the likelihood becomes,
\begin{equation}
    P(d_i|H_1) = \prod_i{\frac{1}{\sqrt{2\pi}\sigma_{d_i}}
        \exp\left(-\frac{(\tilde{d_i}-r_is)^2}{2\sigma_{d_i}^2}\right)}
\end{equation}
where a different standard deviation $\sigma_{d_i}$ is used because the source
contribution adds additional Poisson variation and systematic errror. Assuming
Poisson and systematic errors can be approximated as Gaussian,
\begin{gather}
    \sigma_{d_i}^2 = \sigma_{n_i}^2 + r_is + \sigma_{r_i}^2s^2 \quad (s \ge 0) \\
	    \sigma_{n_i}^2 = \langle n_i \rangle + \sigma_{b_i}^2
\end{gather}
where $\langle n_i \rangle$ is the estimated background, $\sigma_{r_i}^2$
represents systematic error in the model response, and $\sigma_{b_i}^2$
represents systematic error in the estimated background (e.g. fit uncertainty).
The log likelihood-ratio $\ln[P(d_i|H_1) / P(d_i|H_0)]$ captures the relative
support in the data for hypothesis $H_1$ vs $H_0$, and ranks plausible
foreground windows in the GBM data.

Through the prediced counts $r_is$, the likelihood-ratio is dependent on
assumed source amplitude at the Earth $s$, as well as source spectrum, position
on the sky, and Earth position (during the foreground interval) -- all of which influence the model response
$r_i$. A semi-analytic approximate marginalization over source amplitude $s$ is
performed in log-likelihood space, using a power-law prior that favors
directions in which the detector array is more sensitive. 
Marginalization over source
location and spectrum is done numerically, after folding in any potential
location prior. The maximum-liklihood spectrum is also recorded in order to
further classify events. The use of a foreground interval has implicitly
assumed a rectangular light-curve prior with constant spectrum. Marginalization
is not done over foreground interval, instead a down-selected set of
non-overalpping foreground windows with maximum likelihood are saved as event
candidates. Further details are provided in \cite{blackburn2015}.

\section{Significance of two-parameter coincidence\label{app:fap}}

Consider a background of Poisson-distributed events \citep{ross2014introduction} with a particular rate
distribution in threshold parameter $\rho$, thus the density and cumulative
distributions are $d\lambdac / d\rhoth$ and $\lambdac(\rhoth)$ where we have
used subscripts on $\lambdac$ and $\rhoth$ to emphasize the use of cumulative
rate (rate of events with $\rho > \rhoth$).  We would like to calculate the
significance (accidental coincidence probability) of an event from this
population falling within $T$ of a time-of-interest $t_0$. Ordinarily one could
pick in advance a single threshold $\rhoth$ giving a single rate $\lambdac$,
then use the Poisson probability of falling within a certain time window
$P(\Delta t < T) = 1-e^{-\lambdac T} \approx \lambdac T$ for small $P$.  If
many different thresholds are tested, the accidental coincidence probability
may be multiplied by a trials factor representing the different effective
populations of events.

It is convenient to not have to choose a particular threshold in advance, and
thus be able to consider a wide range of possible event rates. In this case,
one must generate a single detection statistic to rank plausible coincidences
between events characterized by the two parameters $\rho$ (or equivalently
$\lambdac$) and $T$ (closeness to $t_0$). A natural ordering is by inverse
false-alarm probability $R = (\lambdac T)^{-1}$. $\lambdac T$ was the original
accidental coincidence probability $P$ for a single threshold, but in this case
we must add up contributions to accidental coincidence from all possible
combinations of $\lambdac$ and $T$ in order to get a faithful representation of
the probability of a coincidence happening with greater $R$ than our event
under consideration. We can calculate the expected number of more highly-ranked
events,
\begin{equation}
    N(R > 1/\lambdac T) = \int_0^\infty d\lambda \int_0^{\lambdac T / \lambda} dt\, e^{-t\,d\lambda}
\end{equation}

By representing the calculation as a sum over slivers of $d\lambda$, we can
conveniently bypass details about the actual shape of $\lambdac(\rhoth)$. Each
sliver actually has the same Poisson distribution $d\lambda\,e^{-t\,d\lambda} =
d\lambda + O(d\lambda^2)$ since they all cover the same amount of differential
rate. However the order itself is determined by the cumulative rate, which sets
the limit of integration. The exponential reduces to first-order in
infinitesimal $d\lambda$ (flat) and the integral becomes,
\begin{equation}
    N(R > 1/\lambdac T) = \int_\lambdamin^\lambdamax d\lambda \, \frac{\lambdac T}{\lambda} = \lambdac T \ln\left(\frac{\lambdamax}{\lambdamin}\right)
\end{equation}
Where $\lambdamax$ and $\lambdamin$ are necessary for convergence.

$\lambdamax$ is naturally constrained by the production threshold of the
events, or by the minimum measurable coincidence time $\lambdamax \Tmin =
\lambdac T$. We can also choose a maximum coincidence window $\Tmax$, up to the
live-time of the experiment, to set $\lambdamin\Tmax = \lambdac T$. Events from
$0 < \lambda < \lambdamin$ will still contribute to the accidental coincidence
probability but subject to a bounded interval of time $\Tmax$. Therefore we
need to add a constant to the expectation value equal to $\lambdac T$. Under
these constraints the expected number becomes,
\begin{align}
    N(R > 1/\lambdac T) &= \lambdac T \left[1+\ln\left(\frac{\lambdamax\Tmax}{\lambdac T}\right)\right] \quad \text{or} \\
    N(R > 1/\lambdac T) &= \lambdac T \left[1+\ln\left(\frac{\Tmax}{\Tmin}\right)\right]\label{eq:fap}
\end{align}
depending on choice of using $\lambdamax$ or $\Tmin$. A two-sided coincidence
window will multiply $N$ by a trials factor of two.  The accidental coincidence
probability $P \approx N$ for small $N$.

\section{Detector data for GW150914-GBM\label{sec:lc}}

\begin{figure}
  \centering
   \includegraphics[width=7in]{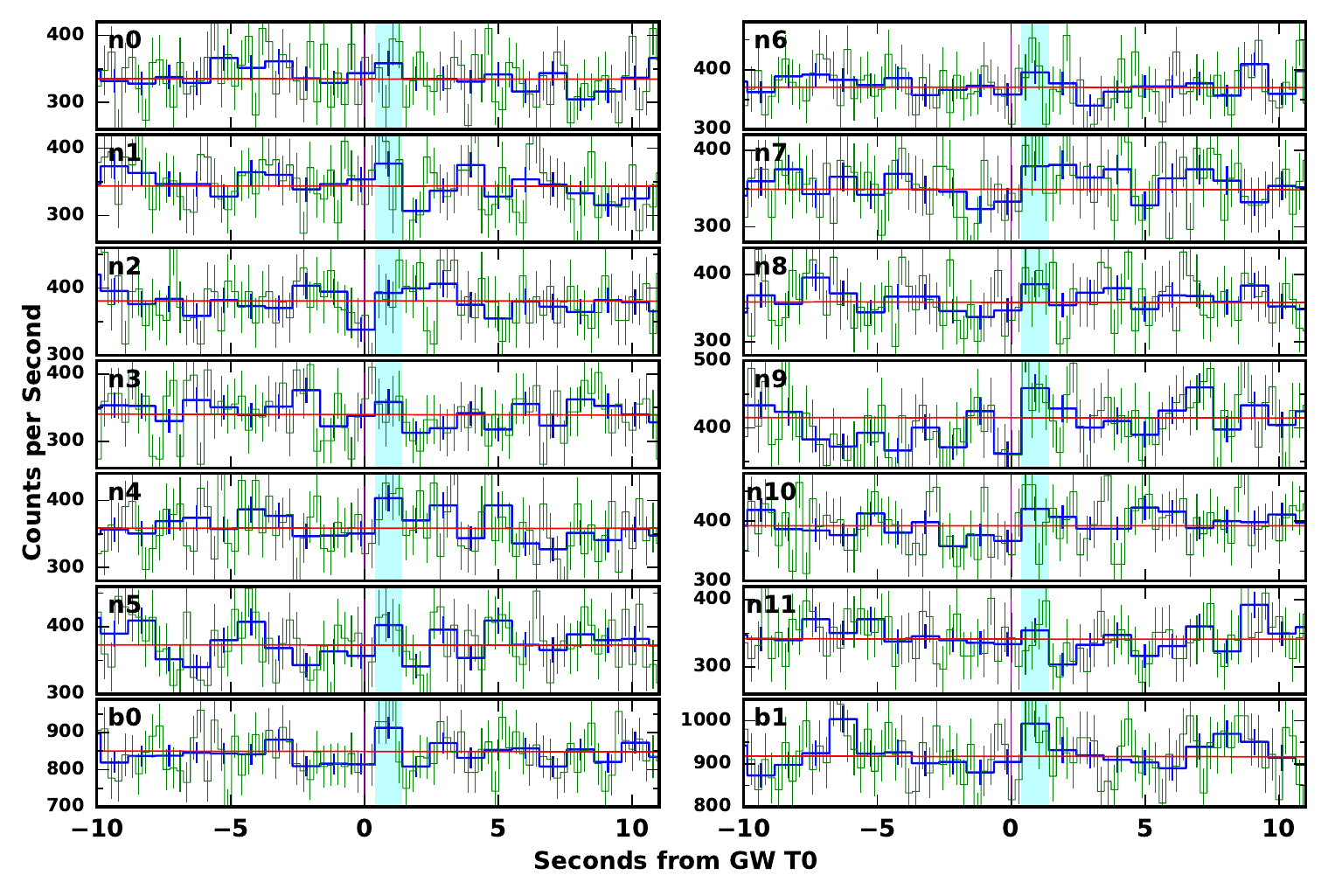}
   \caption{Count rates detected as a function of time relative to the detection time of GW150914,
in each of the 14 GBM detectors.
The shaded region is the time interval of GW150914-GBM, beginning 0.384~s after GW150914,
at 09:30:45.775~UT. 
Time bins are 1.024~s wide and the red line indicates the background.
The blue lightcurve was constructed from CTTE data, rebinned to optimize the
signal-to-noise ratio. 
The 0.256~s CTIME binning is overplotted on the 1.024~s lightcurve.
 NaI data are summed over 50 -- 980 keV and BGO data over 420 keV -- 4.7 MeV.
The detector angles to 
different sky positions on the LIGO localization map are given in Table \ref{tab:arc}.
\label{fig:lc}}
\end{figure}

\begin{figure}
\centering
   \includegraphics[width=4in]{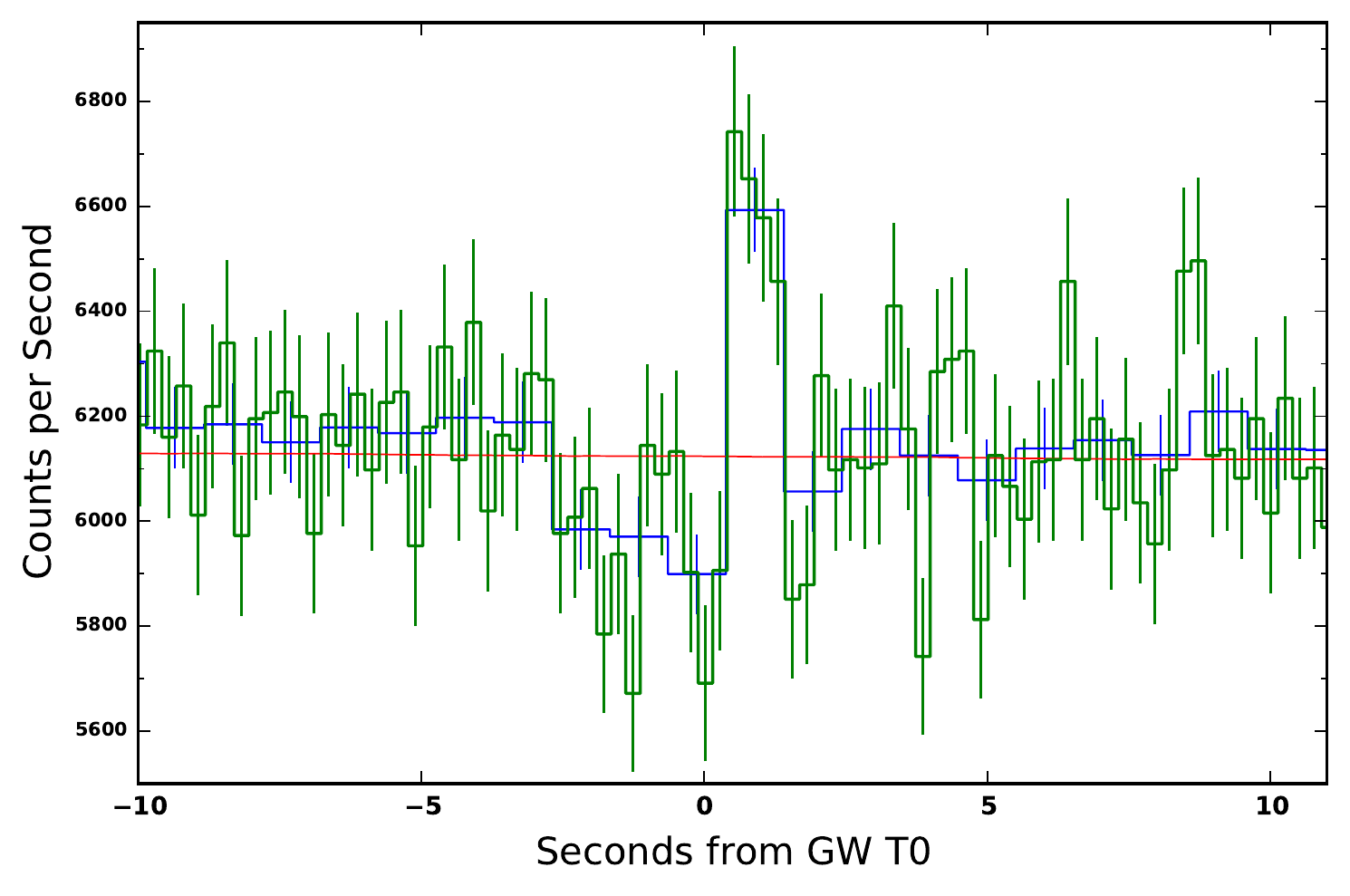}

   \includegraphics[width=4in]{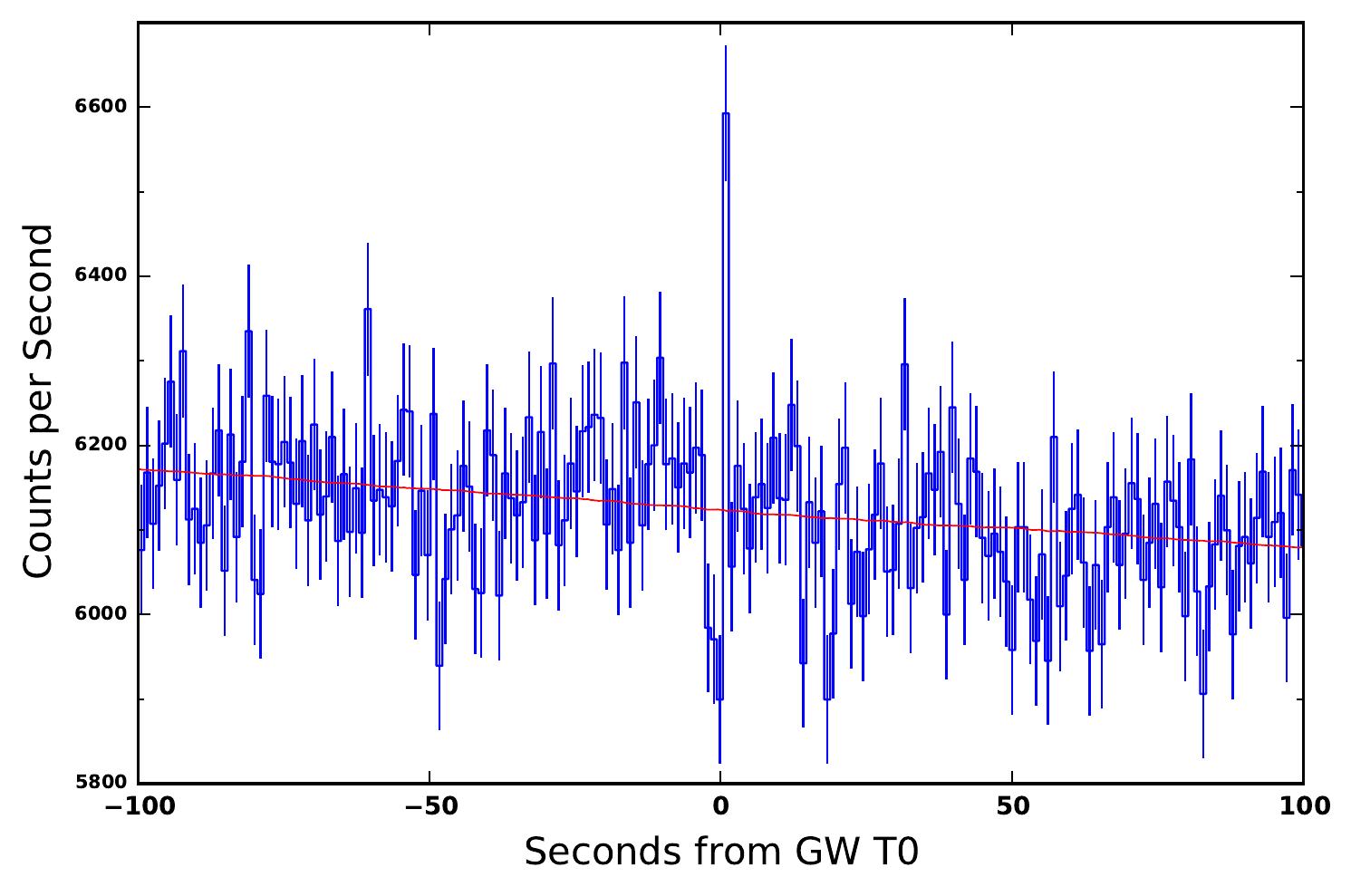}
   \caption{Count rates detected as a function of time relative to the detection time of GW150914, summed over all 14 GBM detectors.
NaI data are summed over 50 - 980~keV and BGO data over 420 keV -- 4.7~MeV.
Time bins are 1.024~s wide, with the
same time binning as in Figure \ref{fig:lc}, and the red line indicates the background level.
The blue lightcurve was constructed from CTTE data, rebinned to optimize the
signal-to-noise ratio.
The signal-to-noise ratio for this
lightcurve, summed over all detectors in the selected energy range, is $6\sigma$. 
In the top panel, the 0.256~s CTIME binning is overplotted on the 1.024~s lightcurve.
The dip before the spike associated with GW150914-GBM is 
not significant.  Such dips are common in stretches of GBM data, as can be seen in the longer stretch of data on the bottom panel.
A 1600~s stretch of data centered on GW150914-GBM, with 1.024~s binning, 
 shows 100 runs each of positive and negative dips lasting 3~s or longer
relative to a third-order polynomial fit background over the 1600~s time interval, with 55 (38) negative (positive) excursions lasting 4~s or longer. 
\label{fig:lc_tot}}
\end{figure}

Figure \ref{fig:lc} shows the count rate registered in all 14 GBM detectors, with a zero time centered on the 
detection time of the GW event.  
In Figure \ref{fig:lc_tot}, the counts are summed over all the detectors.
The time binning of 1.024 s was one of six time-scales (from 0.256 to 8.192~s in multiples of two) selected
{\it a priori} during the optimization of the search procedure, and was the most significant time-scale
over which GW150914-GBM was detected.  
We subsequently optimized the phasing of the 1.024 s bins to produce the largest significance, which is higher
than the significance in the initial 60 s search window (Figure \ref{fig:discovery}). The shaded region shows
this optimized 1.024 s interval, which begins 0.384~s after the GW event, at 09:30:45.775~UT. 

The three low 1.024~s bins in Figure \ref{fig:lc_tot} that precede the high bin are consistent
with a normal background fluctuation. Other similar excursions, positive and
negative, are seen in the panel
showing the longer time span. 
The decrease cannot
be caused by anything blocking photons: for this energy range,
only a very bright and hard transient would be strong enough for
a single source going behind the Earth to cause a rate decrease.   
Nor could a data issue have caused the photons to 
``move'' from the low bins to the high bin that we
attribute to GW150914-GBM, because the GBM hardware time-tags individual photons
as they arrive. There is a known GBM hardware anomaly in which dips and
peaks in a time history are digitally created.  For one second the GBM
clock is mis-set by 0.1 s.  This has the effect of shifting a block of
counts by 0.1 s, leaving a 0.1 s interval with no counts and another 0.1
s interval with double counts -- shifted and correct.  These ``timing
glitches'' are understood and have been extensively studied since they
are readily found by the TGF \citep{briggs2013} and GRB offline
searches.  While there are some variations on this pattern, all timing
glitches are definitively revealed by a time interval of duration tens of
milliseconds with no counts from any detector.  We have examined the
data at higher resolution than shown in Figure \ref{fig:lc_tot}
 and no timing glitches are
present. We have also investigated the
possibility of any telemetry issues or anomalies
suggestive of data problems and we
find that everything on the spacecraft and in our ground processing
was operating nominally.


The lack of a prominent, bright signal in a detector or pair of detectors accounts for the non-detection
of this event on-board and in the undirected offline search.
None of the detectors reaches the single-detector threshold of the offline
search, indicating an event much weaker than the limiting
sensitivity of the undirected search.  The fact that all the NaI detectors, and both BGO detectors, 
register counts above the background fit is unusual.  
In an {\it ad hoc} experiment to quantify how unusual it is,
we looked through 30 days (1.7 million seconds of livetime)
of data for similar features showing high multiplicities of detectors
above or below the background level. 
The signature required
both BGOs to exceed background by  $\geq 2\sigma$, at least two NaI detectors
with $\geq 2\sigma$, and at least six additional
NaI detectors with signal levels $\geq 1\sigma$, for a total
of eight NaI detectors and two BGO detectors with signal requirements.  
Three timescales of the 1.024~s binned data: 
0.7~s, 1.0~s, and 1.4~s, were searched using four 
search window phases and five energy ranges, including
those in the lightcurve shown in Figure \ref{fig:lc_tot}.

GW150914-GBM exceeds these requirements (Table \ref{tab:multi}),
with two NaI detectors above $2\sigma$ and eight additional NaI detectors
above $1\sigma$. 
The search found 20 candidates (including GW150914-GBM), 
14 excesses, and 6 deficits, giving
a 90\% confidence level upper limit of 27.8 total candidates.
If we consider these candidates to be non-astrophysical, this suggests a background
rate of one per $6.12 \times 10^{4}$~s 
implying a chance coincidence of $1.0 \times 10^{-3}$  for a signal  
to accidentally match the signature of GW150914-GBM in a 60~s
period. 

\raggedright

\normalsize


\begin{table}[!h]
\caption{Signals in the GBM detectors in $\sigma$ deviation from
a background fit for the 1.024 second 
high bin in Figure \ref{fig:lc_tot}.  \label{tab:multi}}
\vspace{0.2cm}

\begin{tabular}{cccccc}
NaI 0  &  NaI 1  &  NaI 2  &  NaI 3  &  NaI 4  &  NaI 5  \\
1.31   & 1.81    & 0.64    &  1.05    & 2.42    &  1.68   \\ \hline
NaI 6  & NaI 7  &   NaI 8  &  NaI 9  &  NaI 10 &  NaI 11 \\
1.31   & 1.64   & 1.45     & 2.20   & 1.61    &  0.66    \\ \hline
BGO 0  &  BGO 1  \\
2.25   & 2.56  \\
\end{tabular}
\end{table}


\begin{figure}
  \centering
    \includegraphics[width=7in]{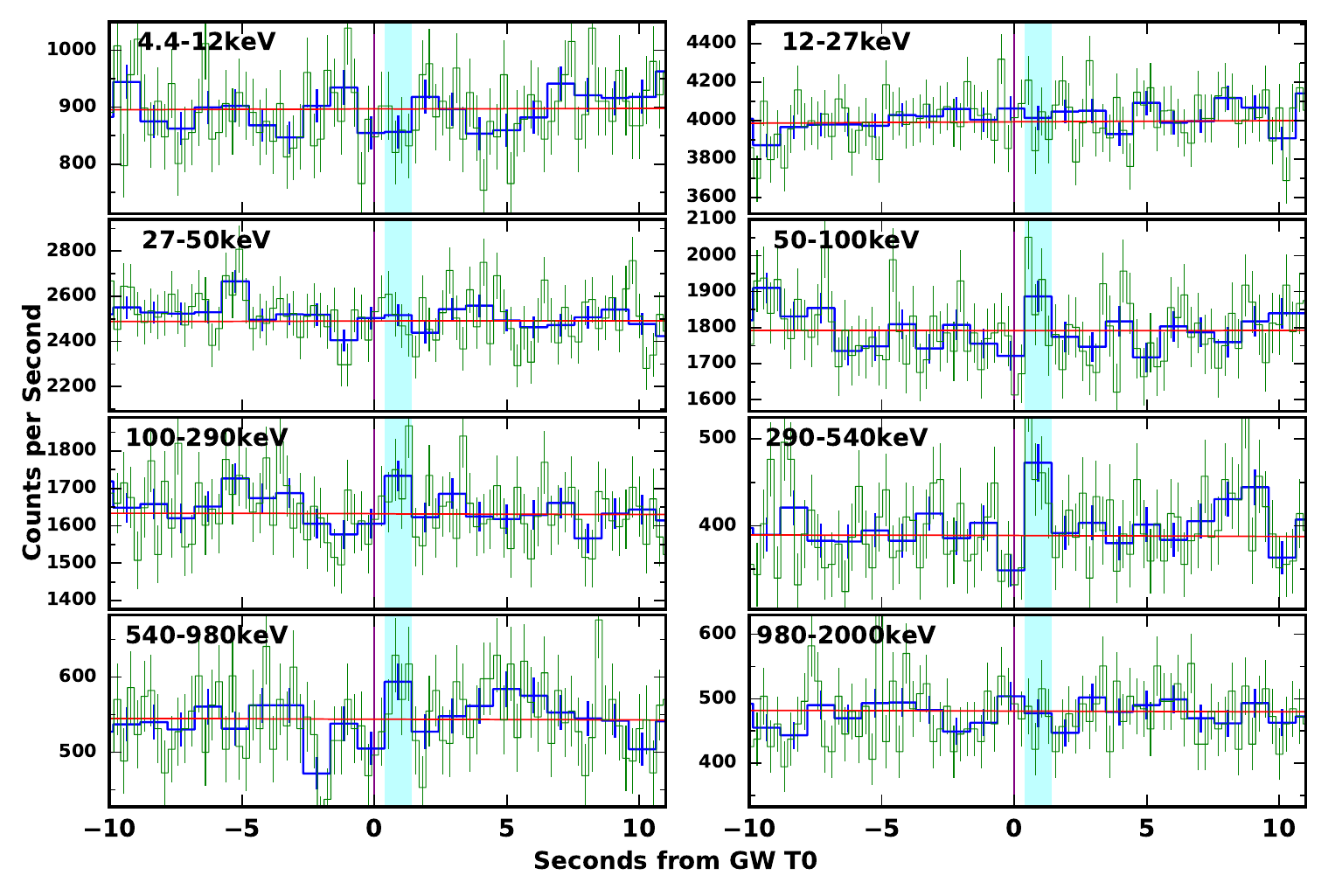}
   \caption{Detected count rates summed over NaI detectors
in 8 energy channels, as a function of time relative to the start of the GW150914. 
Shading highlights the interval containing GW150914-GBM.
Time bins are 1.024~s in duration, with the 0.256~s CTIME lightcurve overplotted in green,
 and the red line indicates the background level. \label{fig:nai_lc}}
\end{figure}

\begin{figure}
  \center{
    \includegraphics[width=7in]{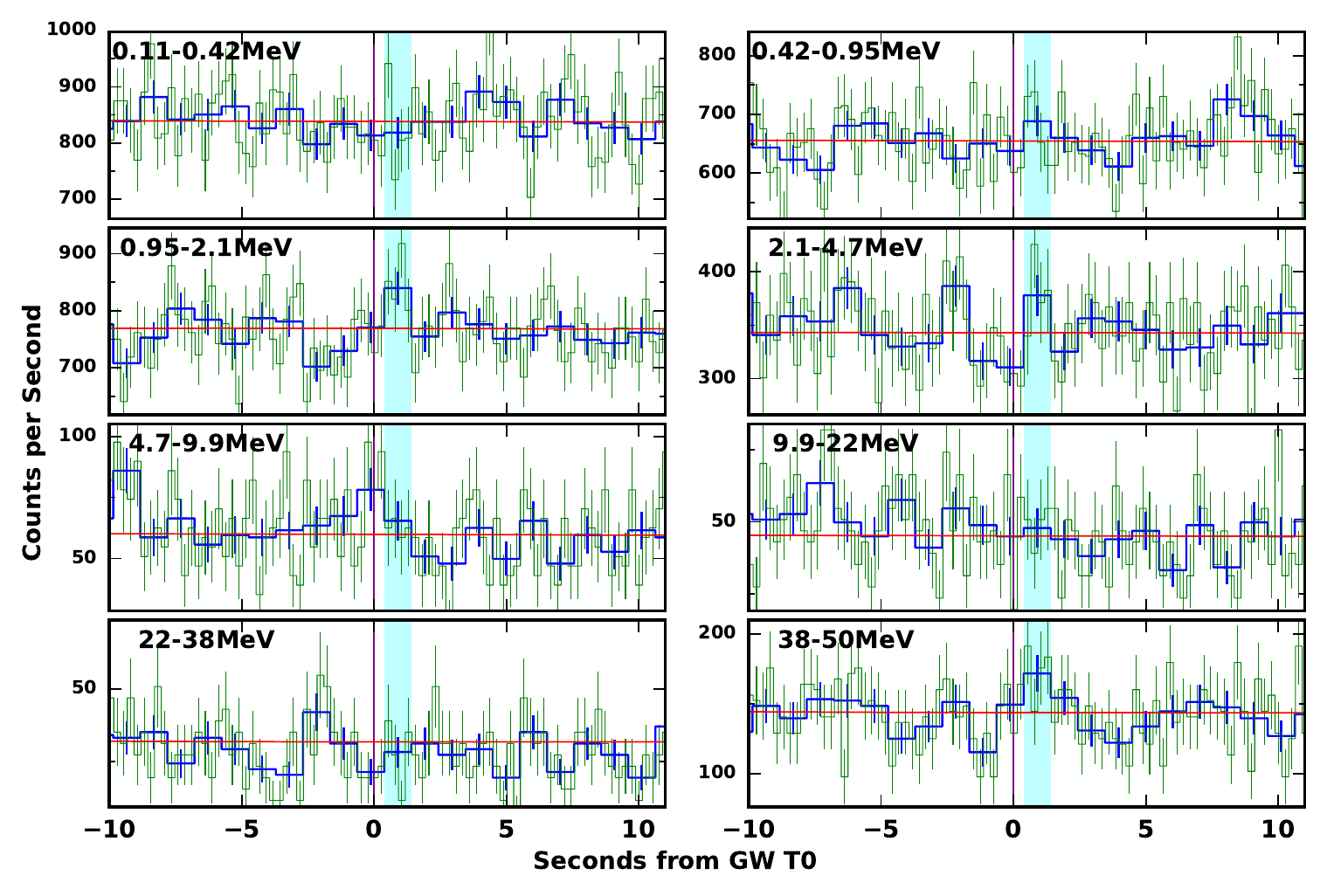}
   \caption{Detected count rates summed over BGO  detectors
in 8 energy channels, as a function of time relative to the start of the GW event. 
Shading highlights the interval containing GW150914-GBM.
Time bins are 1.024~s in duration, with the
0.256~s CTIME lightcurve overplotted in green, and the red line indicates the background level.}
\label{fig:bgo_lc}}
\end{figure}

Figures \ref{fig:nai_lc} and \ref{fig:bgo_lc} shows the lightcurve in the summed
NaI and BGO detectors, respectively, divided into the eight native CTIME energy channels, with the energy ranges indicated
in the panels.   These lightcurves show that GW150914-GBM has a very hard spectrum, with little to no signal below 50 keV
and a peak in the spectrum for the NaI detectors in the 290 -- 540 keV band.  Above 300 keV, photons deposit
little of their energy in the thin NaI detectors so that the measured energy is much lower than the true incident energy.  A
significant count rate in this energy band in the NaI detectors
implies an incident flux of higher-energy photons, consistent with the BGO 
count spectrum that extends into the MeV energy range.  BGO is a higher-Z material and the detectors are thick, so that
incident MeV photons deposit most or all of their energy in the 
scintillator and the measured energy is a good estimate of the incident energy.
Both the NaI and the BGO count spectra look reasonable, with no indications that the event is a statistical fluctuation: there
are no gaps in the spectra between 50 keV and 980 keV for the NaI detectors and between 420 keV and 4.7 MeV in the BGO 
detectors, as one would expect if the event were spurious; the signal increases with energy, peaks, and then decreases, as expected
from a real source; and the NaI and BGO energy spectra are consistent with each other.

\section{Localization of GW150914-GBM\label{sec:loc}}

Source localization involves a comparison of the observed rates in all 12 NaI detectors
with the rates expected from a source at one of 41168 positions that cover all possible arrival
directions in the spacecraft reference frame.

The 50 -- 300 keV energy range is 
the standard selection for source localization, both to minimize
the effect of short time-scale variability contributed by galactic sources such as Sco X-1 (which have steeply falling energy
spectra above 20 keV) and to maximize the counts in the
energy range in which the detector spectral response is very good (response and energy accuracy
fall above 300 keV).  This energy range captures
the peak in the spectral energy distribution for most GRBs.
Model rates are calculated for the detector response to sources
with the three different energy spectra described in section \ref{sec:discovery}.
The most likely arrival direction is the one in which $\chi^2$ is minimized
in a comparison of background-subtracted
observed and model rates on an all-sky grid of $1^\circ$ resolution, as described in \cite{connaughton2015}.
This process yields a localization in both equatorial and galactic coordinates and a 68\% statistical uncertainty radius, $\sigma$.
The uncertainty region covers all the grid points that lie within 2.3 units of the $\chi^2$ minimum, and $\sigma$ is calculated
assuming the uncertainty region is a circle.  In practice the uncertainty region can be irregular in shape
and, for weak events, it may be composed of disjoint islands, so that $\sigma$ is a measure of the size of the uncertainty region
but is not always a good guide to its shape.

\subsection{Standard localization of GW150914-GBM}

The localization of GW150914-GBM finds a best fit to the hard model spectrum and
yields a position of RA, Dec = 57, -22 deg with a 68\% statistical uncertainty region over 9000 square degrees 
($\sigma = 54^\circ$).
 In addition to the large uncertainty, the $\chi^2$ suggests a bad fit to the observed rates that would have
failed a bad-$\chi^2$ cut applied in the regular GBM localization
procedure for GRBs \citep{connaughton2015}. The best-fit location is towards the Earth but the large uncertainty on the location
allows an arrival direction from the sky. 
Figure \ref{fig:lc} shows
that the rates in the NaI detectors are not very high above background and the differences among them do not allow much discrimination of arrival
direction.  GBM detectors register signal counts directly from a source and 
also record a source signal from gamma rays scattering in the Earth's atmosphere, with a magnitude determined
by the source-Earth-detector geometry. When finding the most likely arrival direction for an event,
the localization algorithm fits both a direct and atmospheric component that takes into account the
position of the Earth in the spacecraft coordinate system at the time of the observation. At the time GW150914-GBM
was detected, only one of the NaI detectors had a favorable Earth-viewing angle. The detector normal of NaI 11 was oriented
at $39^\circ$ to the Earth, yet registered the lowest signal above background of 
any detector, suggesting that whatever the source direction, the atmospheric
component was not large.  
NaI detectors 0 through 5 were not susceptible to any flux from the atmosphere because they faced the sky with
the spacecraft positioned between the detectors and the Earth.
There is no weighting in the localization algorithm to disfavor the part of the sky that is occulted by the Earth
 -- the
algorithm uses only the relative rates in the NaI detectors to reconstruct the most likely arrival direction after 
modeling the response to both direct and atmospheric components at each tested sky position (even those behind the Earth),
taking into account the position of the Earth when evaluating the atmospheric component.

Since the detection of GW150914, the analysis of the LIGO data has resulted in
a refinement of the GW event localization, including a new map \citep{gcn_ligo3} that places
most of the probability in the southern portion of the original arc, with only 6\% in a northern sliver of the arc.
Most of the arc lies at a large angle, $\theta$, to the spacecraft zenith, almost entirely under \Fermi. 
Figure \ref{fig:loc_ligo} shows that part of the southern portion of the arc (25\% of the probability) is hidden to \fermi by the Earth. The
rest of the arc lies above the horizon, at low elevation above the Earth to \Fermi.  
We note that for sources at low elevation, the atmospheric component of the signal is low relative to the direct
component \citep{locburst,harmon2002}, compatible with the low count rate observed in NaI 11.
The position RA, Dec = 57, -22 deg returned by our localization procedure is roughly consistent with the LIGO arc. 
Different data interval and background selections of the GBM data used in the localization led in some cases to
localizations at the spacecraft zenith, an indication that the localization process was not converging. 

GBM is a background-limited instrument
and this event is much weaker than any GRB we would normally localize based on either an
on-board or offline detection. The signal to noise ratio in each detector is low and affected by fluctuations in the background rates. 
We reported in \cite{gcn_gbm} that we could not constrain the location of the transient event uncovered in our search. 
We have, since then, 
investigated our data more closely.  

We do not use the BGO detectors in the standard localization process, because their angular response
depends only weakly on the source direction compared to the response of the NaI detectors. 
Also, because the flux from sources detected by GBM
 declines with increasing energy -- and, for GRBs, falls more steeply above 
$E_{\rm peak} \sim$ 100 -- 500 keV -- source signals are usually more intense in the NaI detectors than in the BGO detectors.
For GW150914-GBM, the signals in individual NaI detectors are weak. 
The fact that there is a detectable signal in the BGO detectors suggests that
if the event is real, then for any reasonable source energy spectrum, it arrived from a direction 
preferentially viewed by BGO detectors relative to NaI detectors.
This picture is compatible with a source direction underneath the spacecraft. 

\begin{landscape}
\begin{table}
\footnotesize{
\caption{Sky locations on LIGO localization arc for GW150914 that were visible to GBM at the time of the GW event. 
The first 10 are on the southern lobe, which contains 94\% of the probability.  The positions are $5^\circ$ apart. 
Positions are given in equatorial (Right Ascension and Declination) and spacecraft ($\phi$, $\theta$) frames. The Large Area Telescope (LAT)
 boresight is at spacecraft zenith, $\theta$ = $0^\circ$. Angles to each detector normal are listed for each position. 
The final column shows the \% probability of the LIGO sky map 
contained in a slice of the arc centered on each position.
The 11th position is on the northern lobe, which contains 6\% of the probability of the localization of GW150914.
The positions behind the Earth to \fermi contain 25 \% of the probability and are not listed here.  
The final position listed in the table is the best localization for GW150914-GBM.
All angles are given in degrees. 
\label{tab:arc}}
\vspace{1cm}
\begin{tabular}{l|l|l|l|l|l|l|l|l|l|l|l|l|l|l|l|l|l|l}
\hline
RA & Dec & SC  & SC  & NaI &  &  &  & & & & & & &  & & BGO  &  & Prob.\\ 
 &  &  $\phi$ &  $\theta$ &  0 & 1 & 2 & 3 &  4 &  5 &  6 &  7 &  8 &  9 &  10 &  11 &  0 &  1 & \%\\
\hline
83.98	& -72.85 & 342 & 160  & 144.8  & 122.0 &  83.1 &   117.8 &  76.1  &  71.2  &  161.5 &  142.0 &  97.3   & 149.2 &  103.3 &  108.6 &  70.8  &  109.2 & 12.1 \\
101.99	&  -73.87 & 349 & 156 & 139.9 &  117.1 &  79.2   & 115.2 &  75.4 &   66.5  &  161.6 &  145.5 &  101.3 &  149.4 &  104.1 &  113.4 &  66.1 &   113.9 & 10.0 \\
118.31	& -72.94 & 354&  151 &  134.9 &  112.3 &  75.6 &   112.0 &  74.2 &   61.6  &  159.9 &  148.3 &  105.0 &  149.3  & 105.4 &  118.3 &  61.3  &  118.7 & 10.3 \\
132.04	& -70.44 &	357 & 147 & 129.9 &  107.6  & 72.4  &  108.5 &  72.8  &  56.7  &  157.0  & 150.1 &  108.3 &  149.0 &  106.9 &  123.2 &  56.5  &  123.5 & 11.2 \\
140.85	& -66.63 & 358 & 142 &  125.2 &  103.3 &  69.9  &  104.4 &  70.7 &   51.7 &   153.1  & 150.5 &  110.9 &  148.7 &  109.0 &  128.2 &   51.5  &  128.5 & 10.3 \\
147.53	& -62.51 & 359 & 137 &  120.3 &  98.8 &   67.4   & 100.3 &  68.9 &   46.7 &   148.8 &  150.2 &  113.5 &  147.5 &  110.9  & 133.2 &  46.5  &  133.5 & 7.4 \\
151.18	& -57.97 & 358 & 132 &  115.5 &  94.5 &   65.5 &   96.0  &  66.9 &   41.7  &  144.3 &  148.8 &  115.6  & 146.2 &  113.0 &  138.2 &  41.5 &   138.5 & 5.8 \\
153.363	& -53.091 & 360 & 127 & 111.2 &  90.8 &   64.7 &   91.2  &  64.0 &   37.0  &  139.4 &  145.9 &  116.5 &  145.2 &  115.9 &  142.9 &  36.7  &  143.3 & 3.7 \\
153.933	& -48.239 & 359 & 122 &	106.7 &  87.1 &   64.0 &   86.6  &  61.6 &   32.2  &  134.5 &  142.8 &  117.4 &  143.5 &  118.4 &  147.7 &  31.8   & 148.2 & 1.8 \\
155.331	& -43.208  & 358 & 116 & 102.5 &  83.7 &   64.1 &   81.7  &  58.6 &   27.7  &  129.5 &  138.9 &  117.4 &  141.9 &  121.4 &  152.1 &  27.1   & 152.9 & 2.0 \\
151.172	& -7.256 & 342 & 84 &	75.4  &  66.7  &  76.2  &  45.6 &   39.5  &  21.9  &  93.6 &   105.2  & 105.6 &  124.1 &  141.1 &  157.9 &  18.7 &   161.3  & 4.8 \\
\hline
75. & -73. & 348. & 163. & 147. & 124. & 84. & 120. & 78. & 74. & 162. & 141. & 96. & 148. & 102. & 106. & 73.4 & 106.6 & N/A (GBM) \\
\hline
\end{tabular}}
\end{table}
\end{landscape}

We perform simulations to quantify how well we expect to localize weak signals that come from directions along the
LIGO arc.
We divide the LIGO arc into 11 positions, 10 on the southern portion, one in the north, excluding the parts of the arc that were
occulted to \Fermi.  The positions are listed
in Table \ref{tab:arc}, which shows each 
position in celestial equatorial and spacecraft coordinates, the 
angle to each of the NaI and BGO detectors, and the probability of the LIGO source lying near each position, based on the LIGO location map.
The positions are $\sim 5^\circ$ apart,
comparable to the accuracy with which GBM could localize a weak triggered transient source using the standard localization techniques.
NaI 5 is the only NaI detector with a source angle less than $60^\circ$ for several of the southern lobe positions.
Above an incidence angle of $60^\circ$, the angular response of the NaI detectors drops significantly.
The detectors are, however, not shielded and thus can register counts from any angle, including through the back of the detectors, which
can detect gamma rays or cosmic rays with about 20\% efficiency relative to on-axis particles.

We calculate the expected count rates in each detector between 50 and 300 keV 
using the detector responses for each of the 10 positions along
the southern lobe of the LIGO arc and a normalization based on the observed event signal. 
For each position, we add background rates derived
 from the observed background rate at the time of the detection of GW150914-GBM, and apply Poisson fluctuations to both source and background in
1000 iterations of the 1 s event at each position. 
Using the background-subtracted count rates in each simulated event, 
we assess our ability to localize such a weak source 
using our standard localization process.
The majority of the simulated events are reconstructed near the arc containing the true positions, 
with large uncertainties.
Count rate fluctuations can lead to poor localizations in the wrong part of the sky.  
We note that a significant number of simulated events
(17\%) are placed behind the Earth.  A simulation of the final position in Table \ref{tab:arc} covering the northern lobe of the LIGO
arc places 4\% of the localizations behind the Earth but, unlike the southern lobe, these localizations behind the Earth have consistently large 
$\sigma$ and bad $\chi^2$.
We conclude that the localization of the observed event
GW150914-GBM behind the Earth with a large uncertainty region of 9000 square degrees is not inconsistent with an origin along the
LIGO localization arc, most likely on the southern lobe.

\begin{figure}
  \centering
   \includegraphics[width=6in]{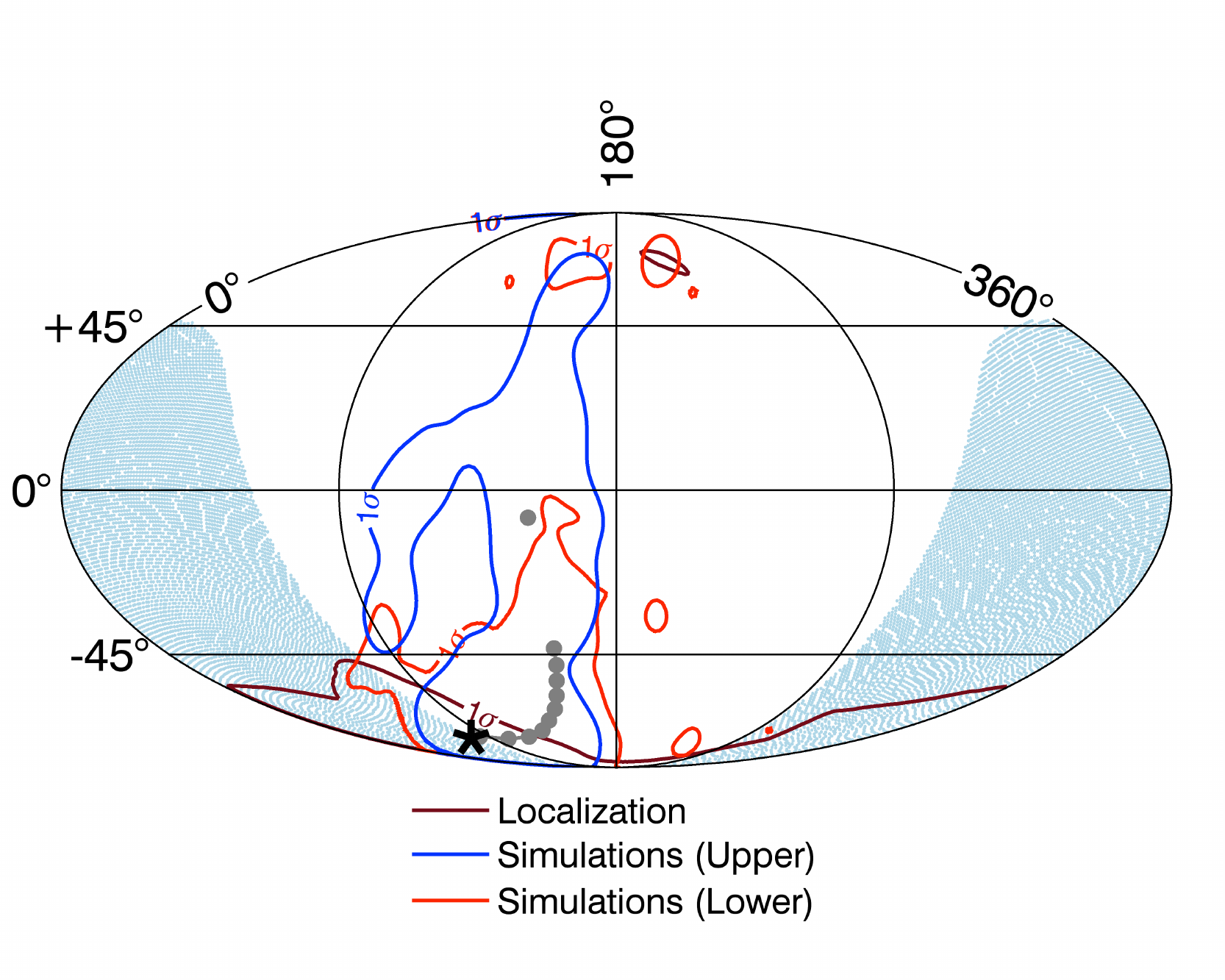}
   \caption{GBM localization of GW150914-GBM using NaI detector counts in
the 100 -- 1000 keV energy range, shown in celestial equatorial coordinates.  The best localization is marked with an asterisk 
and the brown contour indicates the 68\% confidence level region for this localization. 
The best GBM localization is just behind the Earth's limb (shaded blue)
with a large uncertainty contour that significantly overlaps the southern 
lobe of the LIGO location arc (indicated 
as 11 grey circles).   
Simulations in the 100 -- 1000~keV range of the localization of a weak source from each of these 11 positions
along the LIGO localization arc indicate how well GBM localization is expected to perform for a source as
weak as GW150914-GBM with the same source geometry relative to the spacecraft.  The red and blue contours show the 68\% containment for the
simulated locations from the southern (lower) and northern (upper)
 lobe, respectively. The GBM localization overlaps both sets of simulated localizations, with
a better match to those from the southern lobe.
\label{fig:loc_yay}}
\end{figure}

\subsection{Refined localization of GW150914-GBM}

We attempt to refine the GBM localization by examining a broader energy range than the standard 50 -- 300 keV. 
Noting from Figure \ref{fig:nai_lc}
that much of the observed signal occurs above 300 keV, we produced model rates using the soft, medium, and hard spectral models
 in various energy bands, between 50 -- 1000 keV, 50 -- 540 keV, 100 -- 1000 keV, and 100 -- 540 keV. 
We used the standard localization procedure,
minimizing $\chi^2$ for the observed rates in each of the energy ranges relative 
to the model rates in that energy range. 
The localization in each case returned a similar position for the most likely origin of the source, always slightly behind
the Earth, and always at $\theta \sim 160^\circ$. The probability contours 
are more bounded than those from the 50 -- 300 keV localization.  The probability maps cover similar regions of sky for all four localizations.
The smallest statistical uncertainty was found using the 100 -- 1000 keV energy band.
A  minimum was found at RA, Dec = 75, -73 deg with a 68\% confidence region covering about 3000 square degrees 
($\sigma = 30^\circ$) and a preference for the hard spectral model.  The uncertainty contours are broad but constraining.  
With a source this weak from this direction in the spacecraft frame, we reach the limit of being able to use the angular response of the NaI detectors
to localize a source.  The measurement of equal rates in most NaI detectors allows the localization to converge to a region under the spacecraft with
slight discrimination in favor of one or another detector cluster but no further refinement.  We can,
however, say that the general source direction is consistent
with the LIGO arc and define a fairly large region on the sky from which the signal must originate.

We include only statistical uncertainties in our location map.
The standard localization process using the 50 -- 300 keV
energy range was found to have a systematic component on the order of 3 -- 4$^\circ$ for a sample of 200 triggered GRBs \citep{connaughton2015}.  
We do not expect the systematic
error to be much different using the 100 -- 1000 keV energy range, particularly when compared to the size of the statistical uncertainty
when localizing an event this weak, but
we note that our characterization of triggered GRB localizations may not be applicable to these weak events that are more affected by background
fluctuations comparable in size to the signal strength.
Additionally, although the localization uses the standard GBM procedure, the quality of the localizations has not been assessed using non-standard
energy ranges and our uncertainty calculations do not include any systematic component.  

Figure \ref{fig:loc_yay} shows the best position and the associated
1$\sigma$ uncertainty contours for the localization performed using data between 100 and 1000~keV.
The parts of the LIGO arc visible to \fermi
are shown as a series of points (with positions listed in Table \ref{tab:arc}) and the Earth region is shaded.
The LIGO arc overlaps the GBM localization in the southern lobe. 
We also show the 68\% containment region of all the localizations returned by simulations between 100 and 1000~keV
of weak sources from positions
on the southern and northern lobes. 
 The simulations suggest a broad distribution of possible
 locations for a given source position, but we find that the actual localization of GW150914-GBM is quite well constrained to the part of the sky
(and Earth) at high $\theta$, consistent with an origin in the southern lobe of the LIGO annulus.  


\end{document}